\documentclass[a4paper,fleqn]{cas-dc}

\usepackage[switch,mathlines]{lineno}
\let\oldequation\equation
\let\oldendequation\endequation

\renewenvironment{equation}
  {\linenomathNonumbers\oldequation}
  {\oldendequation\endlinenomath}

\usepackage[authoryear]{natbib}

\usepackage{lipsum}

\usepackage{wasysym}

\usepackage{tabularx}
\usepackage{graphicx}
\usepackage{booktabs}
\usepackage{longtable}
\usepackage{caption}
\usepackage{float}
\usepackage{tikz}

\usepackage{siunitx}
\newif\ifarxiv
\arxivtrue

\usepackage{cleveref}

\usepackage{placeins}
\usepackage{flafter}
\usepackage{float}

\usepackage{etoolbox}
\usepackage{xspace}

\newif\ifuseendfloat
\ifuseendfloat
    \usepackage[nolists, nomarkers]{endfloat}
    
\fi

\ExplSyntaxOn
\RenewDocumentCommand \eadauthor {} 
    { 
      \seq_map_inline:Nn \l_stm_au_seq 
        { 
            \regex_extract_once:nnNTF {(\w)\w*-(\w)} { ##1 } \l_stm_au_fn_seq
            { 
                \seq_pop_left:NN \l_stm_au_fn_seq \temp_var
                \seq_use:Nn \l_stm_au_fn_seq { .- }
                { . } 
            }
            { 
                \regex_match:nnTF { \. } { ##1 } 
                { ##1 }
                { \tl_head:n {##1}. }
            }
      }{ ~\l_stm_au_sn_seq }
    }
\ExplSyntaxOff
\newif\ifabbreviation
\pretocmd{\thebibliography}{\abbreviationfalse}{}{}
\AtBeginDocument{\abbreviationtrue}
\DeclareRobustCommand\acroauthor[2]{%
  \ifabbreviation
    \ifcsname acroused@#2\endcsname
      #2%
    \else
      #1%
      ~[\mbox{#2}]
      \expandafter\gdef\csname acroused@#2\endcsname{}%
    \fi
  \else
    \ifcsname bibacroused@#2\endcsname
        \mbox{#2}%
    \else
        \mbox{#1}~(\mbox{#2})%
        \expandafter\gdef\csname bibacroused@#2\endcsname{}%
    \fi
  \fi
}

\usepackage{pdflscape}
\usepackage{rotating}
\usepackage{longtable}

\usepackage{subfigure}

\defcitealias{InternationalOrganizationforStandardization2014ISOFramework}{ISO, 2014} 

\defcitealias{InternationalOrganizationforStandardization2018ISO/TSRequirements}{ISO, 2018} 

\defcitealias{InternationalOrganizationforStandardization2019ISO/TSAnalysis}{ISO, 2019}

\defcitealias{InternationalOrganizationforStandardization2007ISOLanguage}{ISO, 2007}

\defcitealias{iso12913-1}{ISO, 2014}
\defcitealias{iso12913-2}{ISO, 2018}
\defcitealias{iso12913-3}{ISO, 2019}

\newif\ifshowchanges

\definecolor{lightgray}{RGB}{200,200,200}
\ifshowchanges
    \usepackage[commandnameprefix=always]{changes}
    \setcommentmarkup{\marginpar{\small\centering\color{purple}\textbf{[{#1}]}}}
    \setdeletedmarkup{ {\color{lightgray}\sout{#1}}}
\else
    \usepackage[final,commandnameprefix=always]{changes}
\fi

\usepackage{changes}

\begin{document}\sloppy
\let\WriteBookmarks\relax
\def\floatpagepagefraction{1}

\newcommand{\candidate}[1]{\textlangle\textit{#1}\textrangle}
\newcommand{\ccandidate}[2]{\textlangle\textsc{#1}: \textit{#2}\textrangle}

\newcommand{\andcomp}[4]{\ccandidate{#1}{#2}\ and \ccandidate{#3}{#4}}

\newcommand{\ccomp}[4]{\andcomp{#1}{#2}{#3}{#4}}

\newcommand{\pcomp}[2]{\candidate{#1}\ and \candidate{#2}}

\makeatletter
\newcommand{\defcand}[2]{%
  \@namedef{#1}{#2}%
  \@namedef{i#1}{\textit{#2}}%
  \@namedef{c#1}{\candidate{#2}}%
  \@namedef{sg#1}{\ccandidate{sg}{#2}}%
  \@namedef{my#1}{\ccandidate{my}{#2}}%
  \@namedef{insg#1}{\textit{#2} in 
  \textsc{sg}}
  \@namedef{inmy#1}{\textit{#2} in 
  \textsc{my}}
}
\makeatother

\defcand{membi}{membingitkan}
\defcand{menje}{menjengkelkan}

\defcand{meriah}{meriah}

\defcand{menye}{menyenangkan}

\defcand{timer}{tidak meriah}

\defcand{menen}{menenangkan}
\defcand{tenang}{tenang}

\defcand{huru}{huru-hara}
\defcand{kelam}{kelam-kabut}

\defcand{membo}{membosankan}
\defcand{tiber}{tidak berubah oleh itu membosankan}
\defcand{kurang}{kurang kepelbagaian oleh itu membosankan}

\defcand{berse}{bersemarak}
\defcand{rancak}{rancak}

\newcommand*{\papertitle}{Crossing the Linguistic Causeway: A Binational Approach for Translating Soundscape Attributes to Bahasa Melayu}
\shorttitle{\papertitle}    
\shortauthors{B. Lam et al.}  

\title[mode=title]{\papertitle}  
\newif\ifkarnmoved

\author[eee]{Bhan Lam}[orcid=0000-0001-5193-6560]
\ead{blam002@e.ntu.edu.sg}
\corref{c}\cortext[c]{Corresponding author}
\credit{Conceptualization, Methodology, Software, Validation, Formal analysis, Investigation, Project administration, Data Curation, Writing - Original Draft, Writing - Review \& Editing, Visualization, Supervision}

\author[upm]{Julia Chieng}[orcid=0000-0003-1407-3238]
\ead{chiengjulia@upm.edu.my}
\credit{Conceptualization, Methodology, Validation, Investigation, Resources, Writing - Review \& Editing, Project administration}

\ifkarnmoved
\author[eee,gt]{Karn N. Watcharasupat}[orcid=0000-0002-3878-5048]
\else
\author[eee]{Karn N. Watcharasupat}[orcid=0000-0002-3878-5048]
\fi
\ead{karn001@e.ntu.edu.sg}
\credit{Methodology, Software, Validation, Formal analysis, Data Curation, Writing - Original Draft, Writing - Review \& Editing, Visualization}

\author[eee]{Kenneth Ooi}[orcid=0000-0001-5629-6275]
\ead{wooi002@e.ntu.edu.sg}
\credit{Formal analysis, Resources, Writing - Review \& Editing}


\author[eee]{Zhen-Ting Ong}[orcid=0000-0002-1249-4760]
\ead{ztong@ntu.edu.sg}
\credit{Resources, Project administration}

\author[cnu]{Joo Young Hong}[orcid=0000-0002-0109-5975]
\ead{jyhong@cnu.ac.kr}
\credit{Resources, Funding acquisition}

\author[eee]{Woon-Seng Gan}[orcid=0000-0002-7143-1823]
\ead{ewsgan@ntu.edu.sg}
\credit{Resources, Writing - Review \& Editing, Supervision, Funding acquisition}

\affiliation[eee]{
    organization={
        School of Electrical and Electronic Engineering, 
        Nanyang Technological University%
    },
    addressline={50 Nanyang Ave, S2-B4a-03}, 
    postcode={639798}, 
    country={Singapore}
}

\affiliation[upm]{
    organization={
        Department of Music,
        Faculty of Human Ecology,
        Universiti Putra Malaysia%
    },
    addressline={43400 UPM Serdang}, 
    state={Selangor Darul Ehsan},
    country={Malaysia}
}

\ifkarnmoved
\affiliation[gt]{
    organization={
        Center for Music Technology,
        Georgia Institute of Technology%
    },
    addressline={J. Allen Couch Building, 840 McMillan St NW}, 
    city={Atlanta},
    postcode={30318}, 
    state={GA},
    country={USA}
}
\fi

\affiliation[cnu]{
    organization={
        Department of Architectural Engineering,
        Chungnam National University%
    },
    addressline={34134}, 
    city={Daejeon},
    country={Republic of Korea}
}
\tnotemark[1]
\tnotetext[1]{The research protocols used in this research were approved by the respective institutional review board of Nanyang Technological University (NTU), Singapore [IRB-2021-293] and Universiti Putra Malaysia (UPM), Malaysia [JKEUPM-2019-452].}

\begin{abstract}
Translation of perceptual descriptors such as the perceived affective quality attributes in the soundscape standard (ISO/TS 12913-2:2018) is an inherently intricate task, especially if the target language is used in multiple countries. Despite geographical proximity and a shared language of Bahasa Melayu (Standard Malay), differences in culture and language education policies between Singapore and Malaysia could invoke peculiarities in the affective appraisal of sounds. To generate provisional translations of the eight perceived affective attributes --- \textit{eventful}, \textit{vibrant}, \textit{pleasant}, \textit{calm}, \textit{uneventful}, \textit{monotonous}, \textit{annoying}, and \textit{chaotic} --- into Bahasa Melayu that is applicable in both Singapore and Malaysia, a binational expert-led approach supplemented by a quantitative evaluation framework was adopted. A set of preliminary translation candidates were developed via a four-stage process, firstly by a qualified translator, which was then vetted by linguistics experts, followed by examination via an experiential evaluation, and finally reviewed by the core research team. A total of 66 participants were then recruited cross-nationally to quantitatively evaluate the preliminary translation candidates. Of the eight attributes, cross-national differences were observed only in the translation of \textit{annoying}. For instance, \imenje{} was found to be significantly less understood in Singapore than in Malaysia, as well as less understandable than \imembi{} within Singapore. Results of the quantitative evaluation also revealed the imperfect nature of foreign language translations for perceptual descriptors, which suggests a possibility for exploring corrective measures.
\end{abstract}


\maketitle

\section{Introduction}\label{sec:intro}

The Standard Malay (ISO 639-3: \textsc{zsm}), or Bahasa Melayu, is the national language of Malaysia, Singapore, and Brunei. For Malaysia and Brunei, it is also the sole official language, while it is one of the four official languages of Singapore together with English, Mandarin Chinese, and Tamil. In Brunei, however, Brunei Malay (Bahasa Melayu Brunei; ISO 639-3: \textsc{kxd}) is more commonly used as the \textit{lingua franca} than the Standard Malay \citep{McLellan2016TheDarussalam}. \chreplaced[comment=R1.1]{For consistency and brevity, specific language varieties will henceforth be referenced by their ISO 639-3 language code, i.e. \textsc{zsm} for Bahasa Melayu.}{} 
Geographically, Singapore is located south of Peninsular Malaysia, connected by two short land bridges, the Johor–Singapore Causeway and the Malaysia–Singapore Second Link. Although Malaysia and Singapore have a significant shared history as well as close cultural and linguistic ties to each other, each country has developed separate national identities and cultures. Moreover, both countries have distinct language policies. In Singapore, English is the \textit{lingua franca} and the official working language for government and business. Since 1987, Singapore has adopted a bilingual policy where English is offered as the ``first'' language and ``mother tongue'' as a ``second'' language in all Singapore schools \citep{Alfred1987}. Despite the dominance of English, proficiency in \textsc{zsm} is still relatively high at approximately \SI{90}{\percent} across all age groups within the Malay ethnic community in Singapore \citep{Mathews2020}. In Malaysia, \textsc{zsm} is widely used in education, governmental matters and industries along with frequent use of other languages such as English, Chinese, or Tamil among the bilingual or multilingual community. The importance and prevalence of \textsc{zsm} prevail as a national symbol of unity and identity.
As such, intercultural variations in the acoustico-psychometric properties of the \textsc{zsm} translations may exist across the two populations. 

In fact, the issue of intercultural acoustico-psychometric differences is not unique to \textsc{zsm} --- translation of psychometric terms into any language that is substantially used in more than one cultural and/or ethnic groups may require investigation to ensure intercultural validity. An example of one such language is Portuguese, an official language of ten countries with a combined population of 270 million people. In \citet{Antunes2021}, a bicultural assessment of soundscape attribute translations into Portuguese was performed in Brazil and Portugal, where differences in suitability were observed for secondary translations of \textit{chaotic}. 

Within the Malay language family (Macrolanguage ISO 639-3: \textsc{msa}), another standardized variety called Bahasa Indonesia (ISO 639-3: \textsc{ind}) is also used as an official language in Indonesia. Despite mutual intelligibility with \textsc{zsm}, there exist significant linguistic and cultural differences in the use of the two Bahasa varieties. As such, the translation process of soundscape attributes to \textsc{ind} is considered separately from that of \textsc{zsm} \citep[see][]{Sudarsono2021}.

This work describes ``Stage 1'' of the Soundscape Attributes Translation Project (SATP) \citep{Aletta2020}, whereby 8 perceived affective quality (PAQ) attributes in ``Method A'' of ISO/TS 12913-2:2018 \citepalias{iso12913-2} were translated to \textsc{zsm}. The translation of the PAQ attributes (\textit{eventful}, \textit{vibrant}, \textit{pleasant}, \textit{calm}, \textit{uneventful}, \textit{monotonous}, \textit{annoying}, and \textit{chaotic}) follows a similar process as the Portuguese and Thai translations, whereby a set of preliminary translations were first generated via an expert-led procedure. This is followed by a quantitative evaluation framework introduced in \citet{Watcharasupat2022} to evaluate and identify the most suitable provisional translation(s) for each PAQ attribute. The cross-national differences between the Singaporean and Malaysian populations of the expert-led preliminary \textsc(zsm) translations were examined statistically. \chadded[comment=R1.3]{Essentially, this paper identifies (1) whether there are cross-national differences in \textsc{zsm} translations of the PAQ attributes between Singapore and Malaysia, and (2) potential deviations from the implied circumplexity of the PAQ attributes.}

The remainder of the paper is organized as follows. The data collection method, participant demographics, the quantitative evaluation methodology, and the data analysis approach are described in \Cref{sec:method}. The process of obtaining the preliminary translations is detailed in \Cref{sec:provisional}. Analysis of the quantitative evaluation is described in \Cref{sec:quant}. \Cref{sec:discussion} discusses the implications of the analysis; and \Cref{sec:conclusion} concludes the paper. For brevity, Singapore and Malaysia would henceforth be written in shorthands as \textsc{sg} and \textsc{my}. ISO 639-3: \textsc{eng} and ISO 639-3: \textsc{zsm} would be shortened to \textsc{eng} and \textsc{zsm}, respectively.


\section{Methods}\label{sec:method}
\newcolumntype{Z}{>{\raggedright\arraybackslash}X}
\begin{table*}[]
\caption{Summary of sub-stages in the SATP Stage 1 translation from English to Bahasa Melayu}
\label{tab:stage1summary}
\begin{tabularx}{\textwidth}{llll}
\toprule
Stage&      Method&     Outcomes& 
Personnel/Participants \\ 
\midrule
{1a}&  Semantic Translation& 
Initial translations&  Qualified translator \\
{1b}&   Focus Group Discussion& Exploratory translations& 3 academics (linguists)  \& 1 moderator (researcher)                    \\
{1c}& Experiential Evaluation & Preparatory translations &
4 academics (non-linguists)  \\
{1d}&   Committee Review &    
Preliminary translations & Core research team (NTU \& UPM) \\
{1e} & Quantitative Evaluation& 
Provisional translations& 
66 participants recruited; Evaluation by core research team  \\ 
\bottomrule
\end{tabularx}
\end{table*}

In Stage 1 of the SATP initiative, the eight perceived affective quality attributes were to be independently translated to the native languages of the registered working groups, without restriction on the translation methodologies \citep{Aletta2020}. The methodology employed for SATP Stage 1 translations to Bahasa Melayu is thus detailed in this section.
 
\subsection{Data collection method}

The process of deriving the final translations from English to Bahasa Melayu was divided into five substages: ({1a}) an initial translation, ({1b}) focus group discussion, ({1c}) experiential evaluation, ({1d}) committee review, and ({1e}) a quantitative evaluation, as summarised in \Cref{tab:stage1summary}. These substages are \chadded[comment=R2.1]{team-based} method triangulation used in seeking corroboration in the translations, \chadded[]{which resembles the team-based \textit{Translation, Review, Adjudication, Pre-testing and Documentation} (TRAPD) framework \citep{Harkness2008,Mohler2016}.}

In Stage 1a, an initial semantic translation was formulated by a qualified translator in Singapore, who is natively and bilingually proficient in English and Bahasa Melayu. The entire Method A questionnaire in ISO/TS 12913-2 \citepalias{iso12913-2} was translated to Bahasa Melayu with special emphasis on providing alternative translations to the eight PAQ attributes, as shown in \Cref{tab:inittransPAQ}. This is followed by a focus group discussion (FGD) led by a soundscape researcher from Universiti Putra Malaysia (UPM) together with three UPM linguistics academics to assess the cross-cultural accuracy and comprehension of the Singapore-based Stage {1a} translations (Stage {1b}).

The Stage {1b} exploratory translations from the focus group discussion (FGD) were used in the experiential in-situ evaluation with four non-linguistic UPM academics from non-acoustic fields with native bilingual proficiency in English and Bahasa Melayu. Owing to the then-imposed COVID-19 movement restrictions in Malaysia, the academics evaluated the PAQ of their residential environments with the single most preferred translated attributes in the Stage {1b} translations. Upon completion of the in-situ evaluation, the participants were further provided with the English version and the rest of the Stage {1b} translations for further comments. Any newly suggested translations and comments from the experiential evaluation yielded an updated set of translations (Stage {1c}).

A provisional set of translations (Stage {1d}) was then determined by the core research team \chreplaced[comment=R1.4]{from Nanyang Technological University, Singapore (NTU) and UPM}{(NTU and UPM)}, wherein the translations were selected based on their unanimity and preference across translation sets in Stage {1a} to {1c}. Subsequently, a consensus-cum-bias evaluation was performed quantitatively on the Stage {1d} translations to yield the most ``bias-free'', lay, and accurate translations of the eight PAQ attributes (Stage {1e}). \chadded[comment=R2.8]{Notably, the provisional translations were decided based on data-driven decisions by the core research team through the interpretation of the statistical analysis of Stage 1e.}

\subsection{Participants}
A total of 73 participants were recruited across all five substages (1a to {1e}), of which 33 were recruited in Singapore (excluding the translator in Stage 1a) and 40 in Malaysia. All substages were conducted remotely and formal ethical approvals were sought in the respective institutions, i.e., the UPM Ethics Committee For Research Involving Human Subjects (JKEUPM) for Stage {1b} to {1d} (JKEUPM-2019-452); and both the JKEUPM and the NTU Institutional Review Board (NTU-IRB) for Stage {1e} (JKEUPM-2019-452 and IRB-2021-293).  

In Stage 1a, a local graduate student majoring in applied linguistics at NTU, with native bilingual proficiency in English and Bahasa Melayu, was recruited to provide an initial translation. Subsequently, three linguistics academics from the Department of English, Faculty of Modern Languages and Communication at UPM, who were bilingually fluent in English and Bahasa Melayu, took part in the FGD in Stage {1b}. The three academics were from different ethnic groups, namely Chinese, Malay, and Dusun. Another four non-linguistic and non-acoustic academics from UPM -- three native Malays and one Indian -- were recruited for the experiential evaluation in Stage {1c}. The provisional translations were determined by the consensus of all the co-authors in Stage {1d}.

For the quantitative evaluation survey in Stage {1e}, a total of 66 participants were recruited via email and messaging services. To achieve a cross-national comparison, approximately equal numbers of participants were recruited in Singapore and Malaysia. The reported country of residence were approximately equal, i.e., 33 (\SI{50.00}{\percent}) in Singapore, 30 (\SI{45.45}{\percent}) in Malaysia, 3 (\SI{4.55}{\percent}) elsewhere. Based on the self-reported language proficiency via the Interagency Language Roundtable (ILR) scale, wherein participants declared their ILR proficiency in both \textsc{eng} and \textsc{zsm}, most participants were bilingually proficient. Few participants reported limited working proficiency in Malay (\SI{9.09}{\percent}) and English (\SI{1.52}{\percent}), and none of the participants reported elementary proficiency or worse. Most participants have resided in Malaysia or Singapore for most of their lives, whereby only 11 (\SI{10.00}{\percent}) reported staying outside Singapore and/or Malaysia cumulatively for more than five years. All the participants were from diverse disciplines, where only \SI{15.15}{\percent} were from audio-related disciplines. A detailed breakdown of the participant demographics, grouped by reported countries of residence is shown in \Cref{tab:demograph}.


\newcommand{\apc}[1]{(\tablenum[table-format=2.1]{#1}\%)}
\newcommand{\zpc}{\phantom{\apc{11.1}}}
\newcommand{\anb}[1]{\tablenum[table-format=2]{#1}}

\newcolumntype{Y}{>{\centering\arraybackslash}p{1.6cm}}
\begin{table*}[t]
\caption{Demographic information of the participants in Stage 1e, grouped by the reported country of residence}
\label{tab:demograph}
\centering
\begin{tabularx}{\textwidth}{YZ*{6}{Y}}
\toprule
&&
\multicolumn{2}{c}{Singapore} & 
\multicolumn{2}{c}{Malaysia} & 
\multicolumn{2}{c}{Others} \\
\cmidrule(lr){3-8}
\multicolumn{2}{l}{Current country of residence} & 
\multicolumn{2}{c}{\anb{33} \apc{50.0}} & 
\multicolumn{2}{c}{\anb{30} \apc{45.5}} & 
\multicolumn{2}{c}{\anb{3} \apc{4.5}}   \\ \midrule
\multirow[t]{4}{=}{\raggedright\arraybackslash %
Length of stay outside MY/SG} 
& 0-1 years  
    & \multicolumn{2}{c}{\anb{28} \apc{84.9}}  
    & \multicolumn{2}{c}{\anb{14} \apc{46.7}} 
    & \multicolumn{2}{c}{\anb{0} \zpc{}} \\
& 1-5 years 
    & \multicolumn{2}{c}{\anb{2} \apc{6.1}}  
    & \multicolumn{2}{c}{\anb{10} \apc{33.3}} 
    & \multicolumn{2}{c}{\anb{1} \apc{33.3}}   \\
& 6-10 years
    & \multicolumn{2}{c}{\anb{2} \apc{6.1}}  
    & \multicolumn{2}{c}{\anb{4}  \apc{13.3}} 
    & \multicolumn{2}{c}{\anb{1} \apc{33.3}}   \\
& more than 10 years
    & \multicolumn{2}{c}{\anb{1} \apc{3.0}}  
    & \multicolumn{2}{c}{\anb{2} \apc{6.7}} 
    & \multicolumn{2}{c}{\anb{1} \apc{33.3}}   \\ \midrule
&& zsm & eng & zsm & eng & zsm & eng \\ 
\cmidrule(lr){3-4} \cmidrule(lr){5-6} \cmidrule(lr){7-8} 
\multirow[t]{5}{=}{Language Proficiency (ILR)}
& Native (5)  
    & \anb{15} \apc{45.5} 
    & \anb{13} \apc{39.4}  
    & \anb{17} \apc{56.7} 
    & \anb{6} \apc{20.0}   
    & \anb{0} \zpc{}
    & \anb{1} \apc{33.3}  \\
& Full Prof.\ (4)   
    &  \anb{6} \apc{18.2} 
    & \anb{11} \apc{33.3}
    &  \anb{4} \apc{13.3} 
    & \anb{10} \apc{33.3}
    &  \anb{1}  \apc{33.3}   
    & \anb{2} \apc{66.7}  \\
& Prof.\ Working (3)  
    &  \anb{8} \apc{24.2} 
    &  \anb{8} \apc{24.2}   
    &  \anb{7} \apc{23.3} 
    & \anb{14} \apc{46.7}
    & \anb{2} \apc{66.7}
    & \anb{0} \zpc{}   \\
& Lim.\ Working (2)
    & \anb{4} \apc{12.1}  
    &  \anb{1} \apc{3.0}
    & \anb{2} \apc{6.7}  
    &  \anb{0} \zpc{} 
    &  \anb{0} \zpc{}
    & \anb{0} \zpc{}  \\ \midrule
\multirow[t]{5}{=}{Discipline}
& Audio-related  
    & \multicolumn{2}{c}{\anb{7} \apc{21.2}}  
    & \multicolumn{2}{c}{\anb{3} \apc{10.0}} 
    & \multicolumn{2}{c}{\anb{1} \apc{33.3}}   \\
& Non-audio HASS 
    & \multicolumn{2}{c}{\anb{4} \apc{12.1}}  
    & \multicolumn{2}{c}{\anb{9} \apc{30.0}} 
    & \multicolumn{2}{c}{\anb{1} \apc{33.3}}   \\
& Non-audio Engr.  
    & \multicolumn{2}{c}{\anb{11} \apc{33.3}}  
    & \multicolumn{2}{c}{\anb{5} \apc{16.7}} 
    & \multicolumn{2}{c}{\anb{0} \zpc{}}   \\
& Non-audio Sciences  
    & \multicolumn{2}{c}{\anb{6} \apc{18.2}}  
    & \multicolumn{2}{c}{\anb{2} \apc{6.7}} 
    & \multicolumn{2}{c}{\anb{0} \zpc{}}   \\
& Others
    & \multicolumn{2}{c}{\anb{5} \apc{15.2}}  
    & \multicolumn{2}{c}{\anb{11} \apc{36.7}} 
    & \multicolumn{2}{c}{\anb{1} \apc{33.3}}   \\  
\bottomrule
\end{tabularx}
Audio-related: Audio, Acoustics, Music, and Musicology; HASS: Humanities, Arts, and Social Sciences; Engr.: Engineering
\end{table*}

\subsection{Quantitative evaluation methodology}

\begin{figure}[t]
    \centering
    \includegraphics[width=\columnwidth]{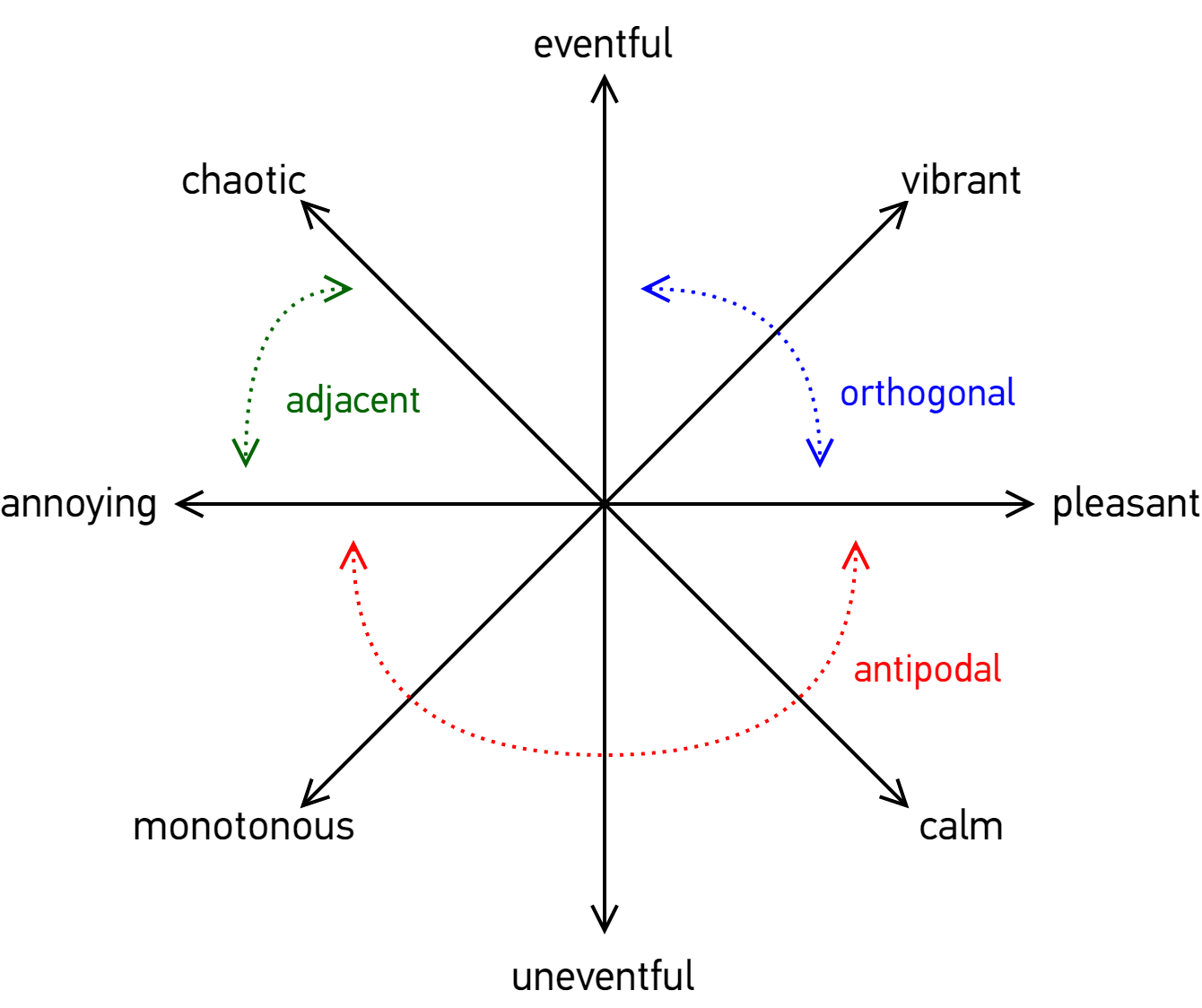}
    \caption{\ifshowchanges\color{blue}\fi Circumplex model of soundscape attributes from ISO/TS 12913-2:2018 \citepalias{iso12913-2} with relational terms introduced in \citet{Watcharasupat2022}. Reproduced from \citet[Figure 1]{Watcharasupat2022}. \ifshowchanges\color{purple}[R1.6]\fi}
    \label{fig:circumplex}
\end{figure}

\chreplaced[comment=R3.1]{The evaluation of the Stage 1d preliminary translations employs}{In} the quantitative evaluation framework proposed by \citet{Watcharasupat2022}, \chadded{wherein} attributes on the main axes (i.e., \textit{eventful}, \textit{pleasant}, \textit{uneventful}, \textit{annoying}) were evaluated on their appropriateness (\textsc{appr}), understandability (\textsc{undr}), clarity (\textsc{clar}), antonymity (\textsc{anto}), orthogonality (\textsc{orth}), non-connotativeness (\textsc{ncon}), and implicative balance (\textsc{ibal}). The attributes on the derived axes (i.e., \textit{vibrant}, \textit{calm}, \textit{monotonous}, \textit{chaotic}) are also assessed on their \textsc{appr}, \textsc{undr}, \textsc{clar}, \textsc{ibal}, as well as on their connotativeness (\textsc{conn}). In the following section, we define the normalized rating of a participant to a questionnaire prompt by $r_{\text{[qn]}}\in[0,1]$ where $\text{[qn]}$ represents the prompt, the contribution of a rating to a score by $s_{\text{[cr]}}\in[0,1]$ where $\text{[cr]}$ represents the criterion. References to adjacency, orthogonality, antipodality of PAQ attributes are based on the octant circumplex model in ISO/TS 12913-3:2019 \citep{iso12913-3}, and depicted in \chreplaced[comment=R1.6]{\protect\Cref{fig:circumplex} \citep[reproduced from][Figure 1]{Watcharasupat2022}.}{Figure 1 of \mbox{\citet{Watcharasupat2022}}.}

The appropriateness score is a direct rating (i.e., $s_{\textsc{appr}}=r_{\textsc{appr}}$) of a translation candidate's suitability in representing the meaning of the English attribute in the context of soundscapes, with a score 0 indicating complete disagreement and a score of 1 indicating complete agreement. 

Understandability is a direct rating (i.e., $s_{\textsc{undr}}=r_{\textsc{undr}}$) in terms of its ease of comprehension by the general population, where a score of 0 represents complete inability in comprehension and 1 represents fully understandable by a layperson.

Clarity measures the degree of association of a candidate with adjacent attributes on the circumplex model of PAQ. It is computed based on penalising the associativeness ratings of the candidate as a translation of clockwise (CW) and counter-clockwise (CCW) adjacent attributes, i.e. $r_{\textsc{asso}}^{\curvearrowleft}$, and  $r_{\textsc{asso}}^{\curvearrowright}$, respectively. Hence, the clarity score is given by
\begin{equation}
    s_\textsc{clar} = 1 - 0.5\left(r_{\textsc{asso}}^{\curvearrowleft} + r_{\textsc{asso}}^{\curvearrowright}\right),
\end{equation}
where a score of 1 implies total clarity and that of 0 implies complete association with both adjacent attributes.

In the circumplex design, antipodal or simply ''opposite" attributes on the main axes should be direct antonyms. Hence, the antonymity score is based on direct rating (i.e., $s_{\textsc{anto}}=r_{\textsc{anto}}$) of the candidate antonymity to the opposite attribute on the main axis.

Additionally, main-axis attributes should be minimally biased towards either CW or CCW adjacent derived-axis attributes. The orthogonality score is based on the deviation of a single biasness rating towards either adjacent attributes (i.e. $r_\textsc{bias}=0$ indicates full biasness towards the CCW attribute and and $r_\textsc{bias}=1$ does so towards the CW attribute). The orthogonality score is given by
\begin{equation}
    s_\textsc{orth} = 1 - 2\left\lvert r_\textsc{bias} - 0.5\right\rvert.
\end{equation}
Hence, $s_\textsc{orth}=0$ indicates full bias towards either of the adjacent attributes and $s_\textsc{orth}=1$ represents complete neutrality (i.e. $r_\textsc{bias} = 0.5$ ). \chadded[comment=R1.5~R1.7]{Since it was previously found that the derived axis attributes were not fully equidistant from the main axes in other languages \citep{Tarlao2016ComparingMontreal,Jeon2018AExperiments,Watcharasupat2022}, and were originally derived from locating attributes that resided along the \SI{45}{\degree} diagonals of the principle components plot near the unit circle \citep{Axelsson2010}, the derived axes are not assessed based on $s_\textsc{orth}$ nor $s_\textsc{anto}$. Instead, the derived attributes should be analysed based on their skew or implicativeness towards the adjacent main axes, which are mostly stable across languages.}

The implicativeness of an attribute towards adjacent attributes are rated via an implicative score for CW ($r_{\textsc{impl}}^{ \curvearrowleft}$) and CCW ($r_{\textsc{impl}}^{ \curvearrowright}$) adjacent attributes, where 0 indicates that the attribute does not imply its adjacent attribute and 1 indicates that it fully implies its adjacent attribute.

Main-axis attributes should not also imply the adjacent attributes on the derived axes, which is represented by a non-connotativeness score written as
\begin{equation}
    s_{\textsc{ncon}} = 1 - 0.5 \left(r_{\textsc{impl}}^{ \curvearrowleft} + r_{\textsc{impl}}^{\curvearrowright}\right).
\end{equation}

In contrast, derived axes attributes should imply both the adjacent main-axis attributes, yielding a connotative score given by
\begin{equation}
    s_{\textsc{conn}} = 0.5 \left(r_{\textsc{impl}}^{ \curvearrowleft} + r_{\textsc{impl}}^{\curvearrowright}\right).
\end{equation}

Lastly, the balance of implicativeness towards either CW or CCW adjacent attributes are determined by the implicative balance score give by
\begin{equation}
    s_\textsc{ibal} = 1 - \left|r_{\textsc{impl}}^{ \curvearrowleft}-r_{\textsc{impl}}^{\curvearrowright}\right|.
\end{equation}
Since translations of the main axes attributes are hardly completely non-connotative, the \textsc{ibal} scores are computed for both main and derived axis attribute translations. 

\subsection{Data analysis} \label{sec:method/dataanalysis}

The cross-national nature of this study resulted in the data being collected in a replicated unbalanced complete block design manner, whereby translation candidates were the blocks to be examined and the countries were the grouping factor. As a result, when testing for inter-country differences, the Prentice test \citep[PT; ][]{Prentice1979} was used for data analysis involving two or more translation candidates under test. The Prentice test is a generalization of the Friedman test  \citep{Friedman1937TheVariance} that can handle replicated, incomplete, and/or unbalanced block design, a situation commonly encountered in cross-population soundscape studies. Where significant differences were found at \SI{5}{\percent} significance level, a pairwise posthoc Mann-Whitney-Wilcoxon rank sum test (MWWT) with Bonferroni corrections was conducted \citep{Bohn1992}. The MWWT null hypothesis was to be rejected at \SI{5}{\percent} significance level. 
For PAQ attributes where only one translation candidate was examined, the Kruskal-Wallis test was employed to examine cross-national differences among the criteria due to unbalanced sample sizes \citep[KWT;][]{Kruskal1952UseAnalysis}. 

For PAQ attributes with multiple translation candidates, the KWT was also employed to examine intra-country differences across all quantitative criteria. 
Where significant differences were found at \SI{5}{\percent} significance level, a pairwise posthoc Conover-Iman test (CIT) with Bonferroni corrections was conducted \citep{Conover1979}. The CIT null hypothesis was to be rejected at \SI{5}{\percent} significance level. 


All data analyses were conducted with the R programming language \citep{RCoreTeam2021} on a 64-bit ARM environment. Specifically, the Prentice test was computed using the package \texttt{muStat} \citep{Wittkowski2012}; Kruskal-Wallis effect sizes were computed with the package \texttt{rstatix} \citep{Kassambara2021}; and the Conover-Iman test was computed using the package \texttt{conover.test} \citep{AlexisDinno2017}. The dataset is available at \hyperlink{https://doi.org/10.21979/N9/0NE37R}{https://doi.org/10.21979/N9/0NE37R} and the code will be made available at \hyperlink{https://github.com/ntudsp/satp-zsm-stage1}{https://github.com/ntudsp/satp-zsm-stage1} 

\section{Preliminary translations: Stages {1a} to 1d}\label{sec:provisional}

Stages 1a to 1d are exploratory investigations into the semantic, experiential, and contextual aspects of the attributes translations. \chadded[comment=R2.2]{Although the binational nature of the translation calls for a validated team-based translation approach \citep{Curtarelli2018,Behr2018}, the TRAPD method was not adopted owing to resource constraints and the nascency of the soundscape field in \textsc{my}. Instead, a modified team-based triangulation approach was designed, which also examines the underlying circumplexity of the PAQ attributes via a quantitative evaluation as the final stage.} \Cref{tab:mainAxisTrans} in Appendix A shows the summary of the translation process of the attributes, including the quantitative evaluation in Stage 1e.


\subsection{Stage 1a: Semantic translation (initial translations)}
In Stage 1a,
\chadded[comment=R2.1-R2.4]{a qualified translator (instead of 2 or more in TRAPD due to cost) who was bilingually fluent in \textsc{eng} and \textsc{zsm} was enlisted in \textsc{sg} to provide an initial list of translations. This was motivated by the preconceived notion that a more appropriate translation could be attained due to greater bilingual proficiency in \textsc{sg}, as well as a higher probability of ensuring a lay translation due to the prevalance of conversational \textsc{zsm} in \textsc{sg} in contrast to formal \textsc{zsm} in \textsc{my}. It is thus of priority to first derive semantic translations from a linguistics perspective for the layperson rather than from a domain perspective. Nonetheless, the translator in Stage 1a was briefed on the purpose of the attributes as descriptors of the acoustic environment, but without detailing the underlying circumplexity of the 8 PAQ attributes.}

An initial translation of five out of the eight attributes generated a single suggestion each, namely \imeriah{} for \textit{eventful}, \imenye{} for \textit{pleasant}, \imembi{} for \textit{annoying}, \imembo{} for \textit{monotonous}, and \textit{kacau-bilau} for \textit{chaotic}. The other attributes have two suggestions each with a preferred choice by the translator: \itimer{} is preferred as compared to \textit{hambar dan senyap} for \textit{uneventful}, \iberse{} as compared to \irancak{} for \textit{vibrant}, and \itenang{} as compared to \textit{tidak kecoh} for \textit{calm}.

\subsection{Stage 1b: Focus group discussion (exploratory translations)}

\chreplaced[comment=R2.1~R2.5 R2.7]{To further evaluate the semantics of the "lay" \textsc{sg}-based Stage 1a translations, a focus group discussion (FGD) consisting of 3 different ethnics, bilingually-fluent linguists from the Department of English, Faculty of Modern Languages and Communication at UPM, was conducted and facilitated by a soundscape researcher at UPM in \textsc{my}. Stage 1b is akin to the TRAPD "review" step, but without the original translator in Stage 1a for a localized (i.e. \textsc{my}) review. Due to pandemic restrictions, the FGD was conducted virtually and the linguists were briefed on the domain specific context of the PAQ translations by the soundscape researcher. Consensus voting was employed to either accept or omit Stage 1a translations, and new translations could also be introduced if deemed appropriate.}{}

\chadded[]{There was unanimous preference in main axis attributes \imeriah{} for \textit{eventful}, \itimer{} for \textit{unpleasant} due to its parallelism with \imeriah{}, and \imenye{} for \textit{pleasant}. In derived axis attributes, \itenang{} for \textit{calm}, and \imembo{} for \textit{monotonous} were unanimously preferred.} \chreplaced[]{For vibrant, \iberse{} was acceptable but \irancak{} appeared more layman, and thus both were retained. In contrast, }{During the focus group discussion with linguistic experts in Stage 1b,}
\textit{hambar dan senyap} for \textit{uneventful} and \textit{tidak kecoh} for \textit{calm}
\chreplaced[]{were unanimously omitted.}{found unsuitable and thus omitted.} 
\chadded[]{The translation \textit{hambar dan senyap} was omitted due to inherent negative connotations although it should be neutral; whereas \textit{tidak kecoh} contains the negation term \textit{tidak}, which deviates from the intended meaning in \textsc{eng}.}

\chadded[]{New translations were also introduced through the FGD, namely, \imenje{} for \textit{annoying} and \textit{kurang kepelbagaian dan membosankan} for \textit{monotonous}. The translation \imenje{} was introduced as \imembi{} is possibly uncommon to the lay person despite being similarly uncommon; whereas \textit{kurang kepelbagaian dan membosankan} was suggested to mean both ``uneventful'' and ``unpleasant'' for greater connotativeness. Hence, a set of exploratory translations consisting of Stage 1a translations after omission (i.e. \textit{hambar dan senyap}, \textit{tidak kecoh}) and addition (i.e. \imenje{}, \textit{kurang kepelbagaian dan membosankan}) of translated attributes were finalised by the lead reviewer (soundscape researcher at UPM) for the next stage.}

\subsection{Stage 1c: Experiential evaluation (preparatory translations)}

\chreplaced[comment=R2.1 R2.6]{A contextual experiential evaluation stage was designed as a local "pretest" with the target population. Hence, four academics who were neither linguists nor acousticians, but were bilingually-fluent in \textsc{eng} and \textsc{zsm} were recruited for Stage 1c. Three of them were native Malays and one was Indian. Due to pandemic movement restrictions, the participants evaluated their surrounding acoustic environments using the \textsc{zsm} questionnaire in Table B.2. Upon completion, the participants were provided with the \textsc{eng} questionnaire and other translations for comments, as shown in Table B.3.}{The contextual translation through experiential evaluation in Stage 1c led to} 

\chadded[]{There was unanimous agreement among the participants for main axis attribute translations. Notably, the translation \imenje{} for \textit{annoying} was shown to the participants but received no comments. For the derived axis attribute translations, translations for \textit{vibrant} were similar with the conclusions in Stage 1b, whereby \iberse{} was less understandable. The translation \itenang{} for \textit{calm} was agreeable by most, and \textit{menenangkan} was additionally suggested to mean ``calming'' by two participants. In translations of \textit{monotonous}, \imembo{} was unanimously preferred as \textit{kurang kepelbagaian dan membosankan} was found to be double-barrelled. An additional translation \textit{tidak berubah} (\textit{lit. unchanging}) was also suggested. In contrast with Stage 1b, there was no consensus in \textit{kacau-bilau} for \textit{chaotic}, and thus \ihuru{} was suggested to describe an atmosphere, and \ikelam{} to describe a place.}

\chadded[]{In summary,} new translations of \textit{menenangkan} for \textit{calm}, \textit{tidak berubah} for \textit{monotonous}, as well as \ihuru{} and \ikelam{} both for \textit{chaotic}, \chadded[]{were suggested in Stage 1c.} 

\subsection{Stage 1d: Committee review (preliminary translations)}

\chreplaced[comment=R2.1 R2.7]{To arrive at a preliminary set of translations, a committee review by the binational core research team was conducted. This is akin to the "adjudication" phase in the TRAPD framework. The candidates were accessed based on consensus voting to form the preliminary list of translations.}{After the research team reviewed the suggested translations and comments in Stage 1d, a list of translations was confirmed.}


The attributes \imeriah{} for \textit{eventful} and \imenye{} for \textit{pleasant} achieved congruence across the four stages of the translation process. For their antitheses \textit{uneventful} and \textit{annoying}, each has two translations. For uneventful, the two translations in Stage 1a are \itimer{} and \textit{hambar dan senyap}. As there is unanimity in the use of \imeriah{} for its opposite attribute of eventful, and \textit{tidak} literally refers to the prefix \textit{``un-''} in uneventful, \itimer{} is preferred for uniformity. For annoying, \imembi{} was agreed as an accurate translation but there was a concern in Stage 1b that it may not be common to the general public. As such, another translation \imenje{} was introduced \chadded[]{in Stage 1b} and both terms were included in \chreplaced[]{the preliminary list}{Stage 1e} to investigate their idiomatic qualities.


The translation process of the attributes in derived axes introduces more alternatives as compared to the translation of the attributes in main axes, possibly due to the need in fulfilling the bilateral condition of each of the four attributes in the derived axes as being \textit{(un)eventful} \textbf{and} \textit{(un)pleasant}. \textit{Vibrant} was translated into \iberse{} and \irancak{} with the latter being considered more of a layperson term. Although \iberse{} is accepted in Stages 1a and 1b, it did not reach a consensus in Stage 1c with one participant viewing it as difficult to understand. To investigate their understandability for soundscape research, both terms are included in the \chadded[]{preliminary list for} quantitative evaluation. \textit{Calm} was translated as \itenang{} and also \imenye{}, with both included \chreplaced[]{in the preliminary list}{for Stage 1e} for further investigation. The other translation for \textit{calm}, \textit{tidak kecoh}, was omitted as the negation \textit{tidak} (\textit{lit.} not) was not used in the original English attribute. 

The attribute \textit{monotonous} is compound in its meaning, i.e., its denotative aspect of being unchanging and its affective quality of being boring. The translation \imembo{} (\textit{lit.} boring) was mostly preferred in Stages 1a to 1c, but there were also other suggestions of \textit{tidak berubah} (\textit{lit.} unchanging) or \textit{kurang kepelbagaian dan membosankan} (\textit{lit.} lack of variety and boring). As boredom can be caused by other factors besides it being uneventful, such as lack of interest, the translation \imembo{} seems to necessitate the attachment of another word that depicts the aspect of it being repetitive, unchanging or lacking variety. Nevertheless, the use of \textit{kurang kepelbagaian dan membosankan} in Stage 1c was commented as double-barrelled and confusing; this translation is perceived as a duality possibly because the two descriptors are connected with the word \textit{dan} (\textit{lit.} and). In English, the twofold meaning of \textit{monotonous} is perceived as one united attribute through the usage of a single word. However, the translation process was unable to propose another one-word literal translation in Malay that contains both psycholinguistic elements. \chadded[comment=R1.8]{The twofold meaning of \textit{monotonous} also posed a problem in other languages as well [i.e. Korean \citep{Jeon2018AExperiments}, French \citep{Tarlao2016ComparingMontreal}].} Hence, another conjunction \textit{oleh itu} (\textit{lit.} thus) was suggested \chreplaced[]{for the preliminary list (Stage 1d)}{in Stage 1d} to form \textit{tidak berubah \textbf{oleh itu} membosankan}, in order to narrow the cause of mundanity down to the unchanging nature of the acoustic environment.

The attribute \textit{chaotic} was initially translated to \textit{kacau-bilau}. There was no disagreement in Stage 1b, but the participants in Stage 1c did not concur with this translation and suggested another two translations of \ihuru{} and \ikelam{}. The contextual usage of each term may be different thus all the latter two translations were included \chadded[]{in the preliminary list} for quantitative evaluations.


\section{Stage 1e: quantitative evaluation (provisional translations)}\label{sec:quant}

\begin{figure*}[t]
    \centering
    \subfigure[]{\includegraphics[width=0.45\linewidth, height=4.5cm]{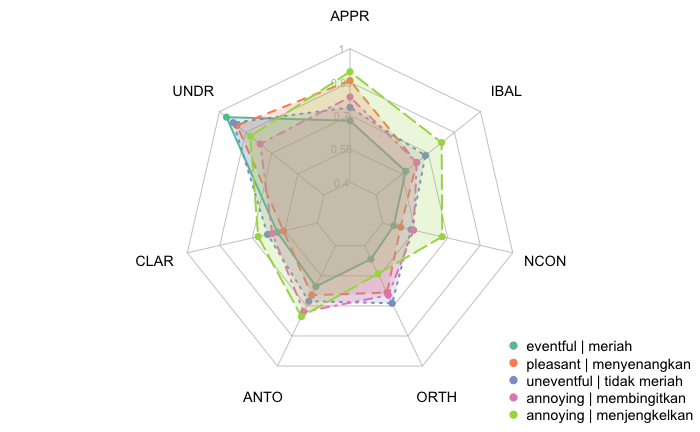}}
    \subfigure[]{\includegraphics[width=0.45\linewidth, height=4.5cm]{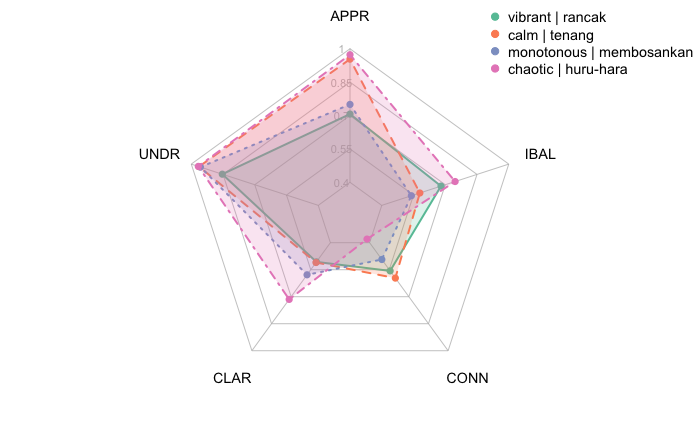}}
    \subfigure[]{\includegraphics[width=0.45\linewidth, height=4.5cm]{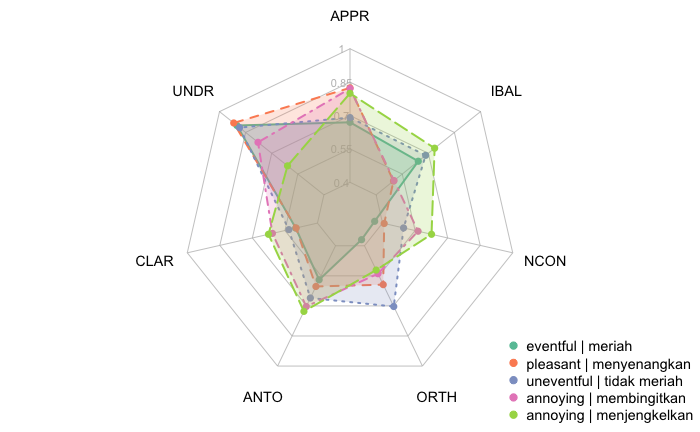}}
    \subfigure[]{\includegraphics[width=0.45\linewidth, height=4.5cm]{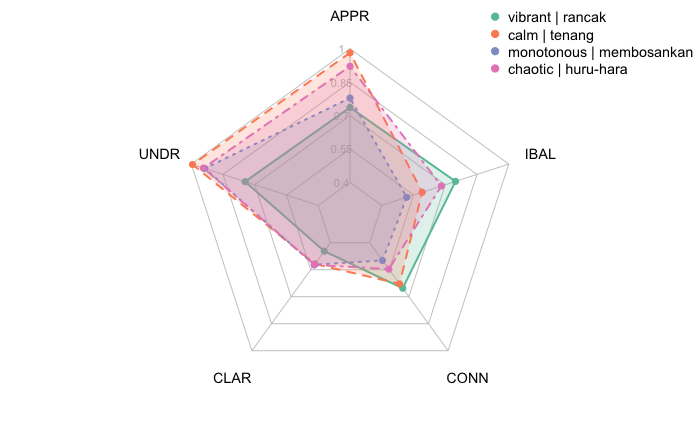}}
    \caption{\ifshowchanges\color{blue}\fi Mean scores of the \textsc{my}-only main (a) and derived axes (b), and \textsc{sg}-only main (c) and derived (d) axes of the participants in stage 1e. Criteria scores of appropriateness (\textsc{appr}), understandability (\textsc{undr}), clarity (\textsc{clar}), antonymity (\textsc{anto}), orthogonality (\textsc{orth}), non-connotativeness (\textsc{ncon}), and implicative balance (\textsc{ibal}) were computed for the provisional translations of main axis; and \textsc{appr}, \textsc{undr},  \textsc{clar}, \textsc{ibal} and connotativeness (\textsc{conn}) for the provisional translations of derived axis attributes. \ifshowchanges\textbf{\color{purple}[R2.9]}\fi} 
    \label{fig:radarFinaltransMYSG}
\end{figure*}

Since the differences between Singapore and Malaysia respondents are being investigated, 3 (\SI{4.55}{\percent}) of the 66 participants whose self-reported current countries of residence (CCR) are outside of Singapore (\textsc{sg}) and Malaysia (\textsc{my}) were excluded in the analyses. Where more than one translation candidate per PAQ attribute were assessed, the differences were evaluated via \chreplaced[comment=R3.2]{PTs,}{Prentice tests\mbox{\citep[{PT;}][{}]{Prentice1979}}}
wherein the candidates and CCR were the blocking and grouping factors, respectively. Only if a statistically significant difference in distributions was found, a posthoc \chreplaced[comment=R3.2]{MWWT with Bonferroni correction}{Mann-Whitney-Wilcoxon rank sum test with Bonferroni correction \mbox{\citep[{MWWT;}][]{RCoreTeam2021}}}
was conducted for pairwise differences between the CCR for each candidate. For cases where only one candidate was assessed, the differences across the CCR split (i.e., \textsc{my} v. \textsc{sg}) were evaluated via the
\chreplaced[comment=R3.2]{KWT}{Kruskal-Wallis test \mbox{\citep[{KWT;}][{}]{Kruskal1952}}.}
The mean criterion scores across CCR are summarised in \Cref{tab:main_scores}\chadded[comment=R2.9]{, and visualized in radar plots in \Cref{fig:radarFinaltransMYSG} for clarity.} Additionally, the $p$-values of the PT and MWWT analyses are summarised in \Cref{tab:prentice} and \Cref{tab:wilcoxon}, respectively. 

\subsection{Main axes} \label{sec:quant_main}

Based on the preliminary translations in Stage 1d, only one translation candidate was selected for \textit{eventful}, \textit{pleasant}, and \textit{uneventful}, i.e., \imeriah{}, \imenye{}, and \itimer{}, respectively, whereas there were two candidates to be evaluated for \textit{annoying} -- \imembi{} and \imenje{}. The main axis translation candidates were evaluated on their \textsc{appr}, \textsc{undr}, \textsc{clar}, \textsc{anto}, \textsc{orth}, \textsc{ncon}, and \textsc{ibal}.

Although inter-candidate comparison for attribute with a single translation cannot be made, the applicability of these candidates were thus evaluated across the CCR via the KWT. No significant differences were found between \textsc{my} and \textsc{sg} for all evaluation criteria (\textsc{appr}, \textsc{undr}, \textsc{clar}, \textsc{anto}, \textsc{orth}, \textsc{ncon}, \textsc{ibal}) on translations of \textit{pleasant} and \textit{uneventful} (i.e., \imenye{}, and \itimer{}, respectively). Hence, translations of \textit{pleasant}, and \textit{uneventful} appear to be suitable across \textsc{my} and \textsc{sg}. For the translation of \textit{eventful} (i.e., \imeriah{}), no significant differences were found across all criteria with the exception of \textsc{undr} ($p<0.05$, $\eta^2<0.06$). This shows that \imeriah{} is significantly more understandable in \textsc{my} ($\mu^{\textit{\textsc{my:}meriah}}_{\textsc{undr}}=0.960$) than in \textsc{sg} ($\mu^{\textit{\textsc{sg:}meriah}}_{\textsc{undr}}=0.900$). 

The differences between the two translation candidates across CCR for \textit{annoying} --- \imembi{} and \imenje{} --- were evaluated via PT on all criteria, where significant differences were found only for \textsc{undr} ($p<0.05$). The posthoc MWWT showed a significant difference between \textsc{my} and \textsc{sg} ($p<0.01$) for \textsc{undr}. This indicates that \menje{} is significantly more understandable in \textsc{my} ($\mu^{\textit{\textsc{my:}menje}}_{\textsc{undr}}=0.823$) than in \textsc{sg} ($\mu^{\textit{\textsc{sg:}menje}}_{\textsc{undr}}=0.609$).

\subsection{Derived Axes} \label{sec:quant_derived}

The translation candidates for derived axis attributes were evaluated on their \textsc{appr}, \textsc{undr}, \textsc{clar}, \textsc{ibal} and \textsc{conn}. Two candidates were evaluated for translations of \textit{vibrant} (i.e., \irancak{}, \iberse{}), \textit{calm} (i.e., \itenang{}, \imenen{}), and \textit{chaotic} (i.e., \ihuru{}, \ikelam{}). For \textit{monotonous}, there were three candidates under evaluation (i.e., \imembo{}, \itiber{}, \ikurang{}). The PT showed no significant differences between translation candidates for \textit{monotonous} across the two CCR groups.

The PT results for \textit{vibrant} revealed significant differences only in \textsc{clar} ($p<0.05$). However, the posthoc MWWT revealed that there were no significant differences between countries for each candidate.

Significant differences were found only in \textsc{conn} ($p<0.05$) in the PT for translations of \textit{calm} across CCR groups. Posthoc pairwise comparisons with MWWT revealed significant differences between CCR ($p<0.05$), wherein \imenen{} had a significantly higher connotative score in \textsc{sg} ($\mu^{\textit{\textsc{sg:}menen}}_{\textsc{conn}}=0.636$) than \textsc{my} ($\mu^{\textit{\textsc{my:}menen}}_{\textsc{conn}}=0.537$).

For \textit{chaotic}, significant differences were found for \textsc{appr} ($p<0.001$), \textsc{clar} ($p<0.001$), and \textsc{conn} ($p<0.05$) in the PT for differences across candidate blocks across CCR groups. The posthoc MWWT revealed significant differences in \textsc{conn} ($p<0.05$) and \textsc{clar} ($p<0.01$) for \ihuru{}. It could be interpreted that \ihuru{} has a significantly higher \textsc{conn} score in \textsc{sg} ($\mu^{\textit{\textsc{sg:}huru}}_{\textsc{conn}}=0.547$) than in \textsc{my} ($\mu^{\textit{\textsc{my:}huru}}_{\textsc{conn}}=0.380$), but has significantly greater clarity in \textsc{my} ($\mu^{\textit{\textsc{my:}huru}}_{\textsc{clar}}=0.715$) than in \textsc{sg} ($\mu^{\textit{\textsc{sg:}huru}}_{\textsc{clar}}=0.523$). It was also found that \ikelam{} was deemed significantly more appropriate in \textsc{my} ($\mu^{\textit{\textsc{my:}kelam}}_{\textsc{appr}}=0.880$) than in \textsc{sg} ($\mu^{\textit{\textsc{sg:}huru}}_{\textsc{clar}}=0.709$).

\section{Discussion}\label{sec:discussion}

\subsection{Cross-national effects on main axis attributes}
Based on the KWT scores in \Cref{sec:quant_main}, no statistically significant differences were found between \textsc{sg} and \textsc{my} across all evaluation criteria for \textit{pleasant} and \textit{uneventful}. Hence, the proposed translations \textit{menyenangkan} and \textit{tidak meriah} are generally applicable across both countries and agrees with the preliminary translations. Interestingly, \textit{menyenangkan} was also independently verified as the most suitable translation for \textit{pleasant} in Bahasa Indonesia \citep{Sudarsono2021}. Even though \textit{meriah} (as a translation of \textit{eventful}) was significantly less understood in \textsc{sg} than \textsc{my}, albeit with weak significance and small effect size, the understandability score of \textit{meriah} is still very high in the \textsc{sg} population ($\mu^{\textsc{sg:}\textit{meriah}}_{\textsc{undr}}=0.9$). Moreover, the \textit{meriah--tidak meriah} translation pair forms a parallelism with the \textit{eventful--uneventful} pair and should thus be used as such. Hence, \textit{menyenangkan}, \textit{meriah}, and \textit{tidak meriah} are confirmed as provisional translations of \textit{pleasant}, \textit{eventful}, and \textit{uneventful}, respectively.

Between \textsc{sg} and \textsc{my} populations, translations for \textit{annoying} were significantly different in \textsc{undr} scores. It was found that \textit{menjengkelkan} was significantly less understandable to the layman in \textsc{sg} than in \textsc{my}. Since 
\SI{95}{\percent} of the Malay-speaking population in \textsc{sg} expressed high levels of proficiency in Bahasa Melayu \citep{Mathews2020}, the lower understandability of \textit{menjengkelkan} could be attributed to its lack of conversational use in \textsc{sg}. Nonetheless, intra-country differences should be examined between candidates before the confirmatory selection of provisional translation candidates.


\subsection{Cross-national effects on derived axis attributes}

Between countries, no significant cross-national differences were found across all criteria for all \textit{monotonous} and \textit{vibrant} translation candidates, as shown in the PT and posthoc MWWT in \Cref{sec:quant_derived}, respectively.

Between both translations of \textit{calm}, cross-national differences were found via PT and MWWT only in \imenen{} in terms of connotativeness. Since the implicative scores towards \textit{pleasant} were similar across countries ($\mu^{\textsc{sg:}\textit{menen}}_{r_{\textsc{impl}}^{\curvearrowleft}}=0.782$, $\mu^{\textsc{my:}\textit{menen}}_{r_{\textsc{impl}}^{\curvearrowleft}}=0.770$), the higher connotative score for \imenen{} in \textsc{sg} could be attributed to a more strongly implied meaning towards \textit{uneventful} ($\mu^{\textsc{sg:}\textit{menen}}_{r_{\textsc{impl}}^{\curvearrowright}}=0.491$; $\mu^{\textsc{my:}\textit{menen}}_{r_{\textsc{impl}}^{\curvearrowright}}=0.303$).

Cross-national differences were also found in translations of \textit{chaotic}, in \textsc{appr}, \textsc{clar}, and \textsc{conn}. The significantly higher connotativeness of \ihuru{} in \textsc{sg} than in \textsc{my} could be attributed to a more strongly implied meaning towards both \textit{annoying} and \textit{eventful} ($\mu^{\textsc{sg:}\textit{huru}}_{r_{\textsc{impl}}^{\curvearrowright}}=0.579$; 
$\mu^{\textsc{my:}\textit{huru}}_{r_{\textsc{impl}}^{\curvearrowright}}=0.380$;
$\mu^{\textsc{sg:}\textit{huru}}_{r_{\textsc{impl}}^{\curvearrowleft}}=0.515$;
$\mu^{\textsc{my:}\textit{huru}}_{r_{\textsc{impl}}^{\curvearrowleft}}=0.380$). Conversely, \ihuru{} has significantly greater clarity in \textsc{my} as a result of lower associations towards \textit{annoying} and \textit{eventful} than in \textsc{sg} ($\mu^{\textsc{sg:}\textit{huru}}_{r_{\textsc{asso}}^{\curvearrowright}}=0.552$; $\mu^{\textsc{my:}\textit{huru}}_{r_{\textsc{asso}}^{\curvearrowright}}=0.317$; $\mu^{\textsc{sg:}\textit{huru}}_{r_{\textsc{asso}}^{\curvearrowleft}}=0.403$; $\mu^{\textsc{my:}\textit{huru}}_{r_{\textsc{asso}}^{\curvearrowleft}}=0.253$). Even though there is a greater desired implied meaning to \textit{annoying} or \textit{eventful}, there is a risk that \ihuru{} could be mistaken as a translation for \textit{annoying} or \textit{eventful} in the \textsc{sg} population. The significantly lower appropriateness of \ikelam{} in \textsc{sg} could be attributed to its uniquely Singaporean colloquial meaning of being ``muddled" or being in a state of mental confusion \citep{Gwee2017}.

\begin{table}[]
\caption{Provisional translations of the 8 ISO 12913-2 PAQ attributes from ISO 639-3: ENG to ISO 639-3: ZSM}
\label{tab:provtrans}
\begin{tabularx}{\columnwidth}{lX}
\toprule
English &
Bahasa Melayu
\\
\midrule                     
pleasant             & {menyenangkan}\\
annoying             & membingitkan (SG); menjengkelkan (MY)\\
eventful             & {meriah} \\
uneventful           & {tidak meriah}\\
vibrant              & {rancak}\\
calm                 & {tenang}\\
monotonous           & {membosankan} \\
chaotic              & {huru-hara} \\

\bottomrule
\end{tabularx}
\end{table} 
 
\begin{figure*}[t]
    \centering
    \includegraphics[width=0.45\linewidth]{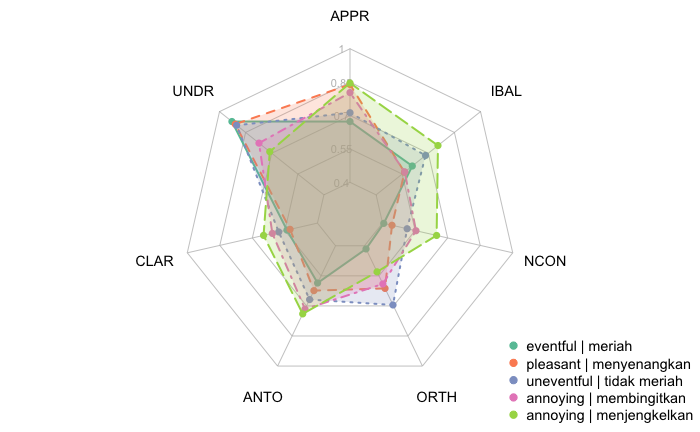}
    \includegraphics[width=0.45\linewidth]{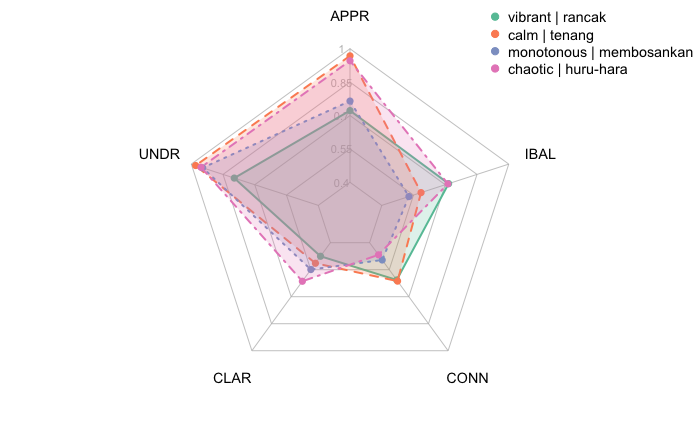}
    \caption{Mean scores of all participants in stage 1e across appropriateness (\textsc{appr}), understandability (\textsc{undr}), clarity(\textsc{clar}), antonymity (\textsc{anto}), orthogonality (\textsc{orth}), non-connotativeness (\textsc{ncon}), and implicative balance (\textsc{ibal}) for the provisional translations of main axis (left); and \textsc{appr}, \textsc{undr},  \textsc{clar}, \textsc{ibal} and connotativeness (\textsc{conn}) for the provisional translations of derived axis attributes (right).} 
    \label{fig:radarFinaltrans}
\end{figure*}

\subsection{Intra-country differences}

As a final step, a KWT was conducted to determine if intra-country differences produced any contradictory results for translations with multiple candidates (i.e., \textit{annoying, vibrant, calm, monotonous, chaotic}).

For the translation of \textit{annoying}, \imembi{} in \textsc{sg} was significantly more understandable ($p<0.01$, $0.06<\eta^2<0.14$), but had a significantly lower \textsc{ibal} score ($p<0.01$, $0.06<\eta^2<0.14$) than \imenje{} in \textsc{sg}. On the contrary, \imenje{} in \textsc{my} was both significantly more appropriate ($p<0.05$, $0.06<\eta^2<0.14$) and had higher non-connotativeness than \imembi{} in \textsc{my}. The intra-country differences suggest that \imembi{} should be preferred over \imenje{} in \textsc{sg} on the basis of understandability, whereas \imenje{} has emerged as the most appropriate translation for \textit{annoying} in \textsc{my}.

Intra-country differences in the translation of \textit{vibrant} was only observed within \textsc{my}, whereby \textit{rancak} was found to be more understandable than \textit{bersemarak}. Therefore, a singular translation via \textit{rancak} was deemed sufficient.

For the translation of \textit{calm},  \insgtenang{} was found to be significantly more appropriate ($p<0.001$, $\eta^2>0.14$) and understandable ($p<0.001$, $\eta^2>0.14$) than \insgmenen{}. Similarly, \inmytenang{} was significantly more appropriate ($p<0.001$,  $\eta^2>0.14$) and understandable ($p<0.05$, $0.06<\eta^2<0.14$) than \inmymenen{}. Thus, the singular translation in \textit{tenang} is suitable in both \textsc{sg} and \textsc{my}.

Across the three candidates for \textit{monotonous}, significant differences were found only in \textsc{undr} for \textsc{sg} ($p<0.01$, $0.06<\eta^2<0.14$), but in both \textsc{appr} ($p<0.05$, $0.06<\eta^2<0.14$) and \textsc{undr} ($p<0.001$, $\eta^2>0.14$) for \textsc{my} in the KWTs. Posthoc CITs revealed that \insgmembo{} was significantly more understandable than both its phrasal variants \insgtiber{} ($p<0.01$) and \insgkurang{} ($p<0.01$). Likewise,  \inmymembo was significantly more understandable than \inmytiber{} ($p<0.001$) and \inmykurang{} ($p<0.05$). Although \inmykurang{} had significantly higher appropriateness over \inmytiber{} ($p<0.05$), it was similarly appropriate to \inmymembo{}. Hence, the singular translation of \textit{membosankan} is thus the most suitable translation for \textit{monotonous}.

Lastly, significant differences were observed only in \textsc{appr} within both \textsc{sg} ($p<0.001$, $\eta^2>0.14$) and \textsc{my} ($p<0.01$,$0.06<\eta^2<0.14$) for the translation of \textit{chaotic}. Since \textit{huru-hara} in both \textsc{sg} and \textsc{my} was significantly more appropriate than \textit{kelam-kabut} in each country respectively, this provides further evidence that \textit{huru-hara} is suitable across and within countries as a translation of \textit{chaotic}.

\subsection{Final provisional translations of PAQ attributes to ISO 639-3: \textsc{zsm}}

Based on the cross- and intra-country analysis of stage 1e, the final set of provisional translations for ISO 639-3:\textsc{zsm} are summarised in \Cref{tab:provtrans}. Except for \textit{annoying}, all other main attributes were represented by the same single-word translation in \textsc{zsm} across both Singapore and Malaysia. For \textit{annoying}, \textit{membingitkan} was selected for \textsc{sg} based on its significantly greater cross- and intra-country understandability, whereas \textit{menjengkelkan} was confirmed as the provisional translation in \textsc{my} due to intra-country advantage in appropriateness and non-connotativeness. 

In the absence of cross-national differences, intra-country understandability was found to be significantly higher in \irancak{} and \imembo{} in \textsc{my} and \textsc{sg}, respectively. Hence, \irancak{} and \imembo{} were confirmed as provisional translations of \textit{vibrant} and \textit{monotonous}, respectively. For translations of \textit{calm}, \itenang{} was significantly more appropriate and understandable than \imenen{} within both \textsc{sg} and \textsc{my}  despite stronger connotativeness of \imenen{} in \textsc{sg} than \textsc{my}. Thus, \itenang{} is selected as the final provisional translation of \textit{calm}. Between the translations of \textit{chaotic}, the significantly lower appropriateness of \ikelam{} in \textsc{sg} over \textsc{my} diminishes its suitability over \ihuru{}. Since intra-country analysis showed that \ihuru{} was significantly more appropriate within both countries, \ihuru{} is thus confirmed as the provisional translation of \textit{chaotic} for both \textsc{sg} and \textsc{my}.

Visualisation of the mean criteria scores of the main axis and derived axis attribute translations in \Cref{fig:radarFinaltrans} also reveals potential incompatibility with the assumed circumplexity of the perceived affective quality model in ISO/TS 12913-3:2019 \citepalias{iso12913-3}. It should be noted that the low \textsc{anto}, \textsc{orth}, \textsc{ncon}, and \textsc{ibal} scores directly affect the integrity of the main axis assumptions, for example in \imeriah{}. Likewise, low \textsc{conn} and \textsc{ibal} scores also affect the structural integrity of the derived axes, for instance in \imembo{}. Hence, there is an impetus to further investigate the circumplexity of the \textsc{zsm} translations as well as means for calibration or compensation for the equations in ISO/TS 12913-3:2019 to be valid. 

\section{Conclusion}\label{sec:conclusion}

This work details the procedure for translating perceptual soundscape attributes in ``Method A'' of the ISO/TS 12913-2:2018 soundscape technical specification \citepalias{iso12913-2} from English (ISO 639-3: \textsc{eng}) to Standard Malay (ISO 639-3: \textsc{zsm}), as a part of Stage 1 in the ``Soundscape Attributes Translation Project'' (SATP) initiative. The methods employed in this study investigated the associated meanings in socio-cultural contexts and applicability in soundscape appraisal of the translations of the eight PAQ attributes into Malay across two countries.

A set of preliminary attribute translations in Standard Malay was first generated by a four-part expert-led process. Even though the expert-led process was a cross-national collaboration between Singapore and Malaysia, it was inadequate to rigorously account for potential cross-national semantic disparities. Hence, a quantitative approach first developed for the Thai language translation in the SATP \citep{Watcharasupat2022} was adopted for the cross-national analysis of the Standard Malay preliminary translations.

The quantitative evaluation framework successfully unveiled cross-national differences in the translation of \textit{annoying}, but otherwise agreed with the preliminary translations. Moreover, analysis of the structural criteria scores across candidates also indicated a potential violation of circumplexity. For instance, the provisional translation \textit{eventful} has a relatively low orthogonality and undesirably high implicative connotations of \textit{vibrant} and \textit{chaotic}, which challenges the circumplexity assumptions and affects the applicability of formula ``A.2'' in ISO/TS 12913-3:2019 \citepalias{iso12913-3}.

Further research can be conducted on the cross-cultural compatibility of the Malay translations among different bilingual or multilingual ethnicities that reside in both countries.

\ifshowchanges \newpage \fi
\section*{Declaration of competing interest}
The authors declare that they have no known competing financial interests or personal relationships that could have appeared to influence the work reported in this paper.

\section*{Acknowledgments}
This work was supported by the National Research Foundation, Singapore, and Ministry of National Development, Singapore under the Cities of Tomorrow R\&D Program (CoT Award: COT-V4-2020-1), and the Google Cloud Research Credits program (GCP205559654). Any opinions, findings and conclusions or recommendations expressed in this material are those of the authors and do not reflect the view of National Research Foundation, Singapore, and Ministry of National Development, Singapore.

The authors would like to thank Dr.\ Francesco Aletta, Dr.\ Tin Oberman, Dr.\ Andrew Mitchell, and Prof.\ Jian Kang, of the UCL Institute for Environmental Design and Engineering, The Bartlett Faculty of the Built Environment, University College London (UCL), London, United Kingdom, for coordinating the SATP project and providing assistance for the Bahasa Melayu Working Group; and Ms. Yun-Ting Lau for her assistance in the data collection.

\section*{Data availability}
The data that support the findings of this study are openly available in NTU research data repository DR-NTU (Data) at \href{https://doi.org/10.21979/N9/AUE2LL}{doi:10.21979/N9/AUE2LL}, and the replication code is available on GitHub at \href{https://github.com/ntudsp/satp-zsm-stage1}{github.com/ntudsp/satp-zsm-stage1}.

\printcredits


\bibliographystyle{cas-model2-names}

\bibliography{refbhan}

\setcounter{section}{0}
\renewcommand{\thesection}{Appendix \Alph{section}}

\setcounter{table}{0}
\renewcommand{\thetable}{\Alph{section}.\arabic{table}}
\onecolumn

\newcommand{\candadded}{\scalebox{0.7}{$\blacktriangleright$}}
\newcommand{\xcandremoved}{\scalebox{0.7}{$\blacksquare$}}
\newcommand{\xcandselected}{\scalebox{1.1}{$\star$}}
\newcommand{\candremoved}{\hfill\xcandremoved{}}
\newcommand{\candselected}{\hfill\xcandselected{}}

\newcommand{\allgood}{unanimous preference}
\newcommand{\alltoquant}{unanimous for inclusion in 1e}

\begin{landscape}

\section{Summary of the Translation Process} \label{sec:appendTrans}

\begin{sffamily}
\begin{nolinenumbers}
\small%
\refstepcounter{table}\label{tab:mainAxisTrans}
\noindent\textbf{\color{scolor}Table \thetable}\par
\noindent{Translation process of the main axis PAQ attributes from English to Bahasa Melayu across all sub stages (1a to 1e)}
\end{nolinenumbers}
\small
\setlength{\tabcolsep}{3pt}
\renewcommand\arraystretch{0.85}
\centering
\begin{tabularx}{\linewidth}{%
l   
>{\raggedright\arraybackslash}p{2.8cm} 
l   
>{\raggedright\arraybackslash}p{5.5cm} 
>{\raggedright\arraybackslash}p{3.9cm} 
>{\raggedright\arraybackslash}X 
l 
}
\toprule
Attribute & 
Translation & 
Stage 1a&
Stage 1b&
Stage 1c&
Stage 1d&
Stage 1e\\
\midrule 
eventful 
    & meriah  
        & \candadded{} sole translation 
        & \allgood{} 
        & \allgood{} 
        & \allgood{} 
        & selected \candselected{}
        \\  \midrule
pleasant 
    & menyenangkan  
        & \candadded{} sole translation
        & \allgood{} 
        & unanimous; accurate reverse translation by a participant & \allgood{} 
        & selected \candselected{}
        \\  \midrule
uneventful 
    & tidak meriah  
        & \candadded{} preferred 
        & preferred in parallel with ``meriah'', as in \textsc{eng} i.e., ``uneventful'' and   ``eventful'' 
        & agreed by most
        & \allgood{} 
        & selected \candselected{}
        \\    \cmidrule(l){2-7}
    & hambar dan senyap
        &  \candadded{} introduced
        &  unanimous omission (negative connotation as ``uneventful'' is neutral)~\candremoved{} 
        & --- 
        & --- 
        & ---    %
        \\  \midrule
annoying 
    & membingitkan 
        & \candadded{} sole translation
        & accurate but may be uncommon 
        & \allgood{} 
        & \allgood{} 
        & selected \candselected{}
        \\ \cmidrule(l){2-7}
    & menjengkelkan
        & ---
        & \candadded{} may be uncommon 
        & no comments 
        & \alltoquant{} 
        & selected \candselected{}\\
            
            

\midrule 
vibrant
    & bersemarak  
        & \candadded{} preferred 
        & can be accepted 
        & difficult to understand 
        & \alltoquant{} 
        & omitted \candremoved{}
        \\ \cmidrule(l){2-7}
    & rancak  
        & \candadded{} introduced
        & more layman 
        & preferred by most 
        & \allgood{} 
        & selected \candselected{}
        \\  \midrule
calm 
    & tenang  
        & \candadded{} preferred 
        & \allgood{} 
        & preferred by most 
        & \allgood{} 
        & selected \candselected{}
        \\ \cmidrule(l){2-7}
    & tidak kecoh  
        & \candadded{} introduced
        & unanimous omission (``tidak'' means ``not''; negation)\candremoved{}
        & --- 
        & --- 
        & ---
        \\ \cmidrule(l){2-7}
    & menenangkan  
        & ---
        & ---
        & \candadded{} suggested by two participants (means ``calming'') 
        & \alltoquant{} 
        & omitted \candremoved{}
        \\ \midrule
monotonous 
    & membosankan  
        & \candadded{} sole translation
        & \allgood{} 
        & preferred by most 
        & \allgood{} 
        & selected \candselected{}
        \\ \cmidrule(l){2-7}
    & kurang kepelbagaian dan membosankan  
        & ---
        & \candadded{} to translate both the ``uneventful'' and ``unpleasant'' aspects of the attribute & double-barrelled 
        & unanimous omission \candremoved{}
        & ---
        \\ \cmidrule(l){2-7} 
    & tidak berubah  
        & --- 
        & --- 
        & \candadded{} only means ``unchanging'' thus lacking inclusiveness
        & unanimous omission \candremoved{}
        & ---
        \\ \cmidrule(l){2-7}
    & tidak berubah oleh itu membosankan  
        & --- 
        & --- 
        & --- 
        & \candadded{} unanimous addition (to~prevent double-barrelled meaning, ``oleh itu'' which means ``therefore'' connects the element of ``uneventful'' [unchanging] to ``unpleasant'' [boring])
        & omitted \candremoved{} 
        \\ \cmidrule(l){2-7}
    & kurang kepelbagaian oleh itu membosankan  
        & --- 
        & --- 
        & --- 
        & \candadded{} unanimous addition 
        & omitted \candremoved{} 
        \\ \midrule
chaotic 
    & kacau-bilau  
        & \candadded{} sole translation 
        & no disagreement 
        & no consensus 
        & unanimous omission \candremoved{}
        & ---
        \\ \cmidrule(l){2-7}
    & huru-hara 
        & --- 
        & --- 
        & \candadded{} to describe an atmosphere 
        & \allgood{} 
        & selected \candselected{}
        \\ \cmidrule(l){2-7}
    & kelam-kabut 
        & --- 
        & --- 
        & \candadded{} to describe a place 
        & unanimous for inclusion in stage 1e 
        & omitted \candremoved{}
\\ 
\bottomrule
\end{tabularx}
\candadded{}: candidate introduced at this stage. \xcandremoved{}: candidate removed after this stage. \xcandselected{}: candidate selected for Stage 2.
\end{sffamily}
\end{landscape}

\setcounter{table}{1}
\renewcommand{\thetable}{\Alph{section}.\arabic{table}}

\section{Translations of the perceived affective quality questionnaire to Bahasa Melayu}

\newcommand{\bigsq}{\scalebox{1.5}{$\square$}}

\begin{sffamily}
\setcounter{table}{1}
\small
\noindent\textbf{\color{scolor}Table \thetable}\\
\normalfont\sffamily Translation of the perceived affective quality (PAQ) questionnaire from ISO/TS 12913-2:2018 \citepalias[C.3.1.3]{iso12913-3} to Bahasa Melayu, as used in Stage 1a. The upright text shows the Bahasa Melayu version as presented to the participants. The following italicized text in square brackets indicates the corresponding English counterpart from ISO/TS 12913-2:2018.
\label{tab:inittransPAQ}
\small
\setlength{\tabcolsep}{0pt}
\vskip1em\noindent
\begin{tabularx}{\textwidth}{%
    l
    @{\hspace{-2.5em}}*{5}{>{\centering\arraybackslash}X}%
}
\toprule
    \multicolumn{6}{l}{%
        Kualiti Afektif Yang Dirasakan [\textit{Perceived Affective Quality}] 
    } \\
    \multicolumn{6}{p{17cm}}{%
        Untuk setiap skala di bawah, sejauh mana anda bersetuju atau tidak bersetuju, bahawa bunyian yang mengelilingi persekitaran anda sekarang adalah... (Sila tandakan satu kotak di setiap baris) \newline
        [\textit{For each scale below, to what extent do you agree or disagree, that the present surrounding sound environment is ... (Please tick off one response alternative per scale.)}] 
    } \\
\midrule
    & Sangat bersetuju             
    & Bersetuju              
    & Bukan tidak~bersetuju mahupun bersetuju    
    & Tidak bersetuju         
    & Sangat tidak~bersetuju \\ 
    & [\textit{Strongly agree}]                      
    & [\textit{Agree}]              
    & [\textit{Neither agree,\newline nor disagree}]       
    & [\textit{Disagree}]              
    & [\textit{Strongly disagree}]   \\ 
\midrule
    Menyenangkan & \bigsq & \bigsq & \bigsq & \bigsq & \bigsq\\
    {}[\textit{Stage 1a trans. ``pleasant''}]\\
    Kacau-bilau & \bigsq & \bigsq & \bigsq & \bigsq & \bigsq\\
    {}[\textit{Stage 1a trans. ``chaotic''}]\\
    Bersemarak; Rancak & \bigsq & \bigsq & \bigsq & \bigsq & \bigsq\\
    {}[\textit{Stage 1a trans. ``vibrant''}]\\
    Tidak meriah; Hambar dan senyap & \bigsq & \bigsq & \bigsq & \bigsq & \bigsq\\
    {}[\textit{Stage 1a trans. ``uneventful''}]\\
    Tenang; Tidak kecoh& \bigsq & \bigsq & \bigsq & \bigsq & \bigsq \\
    {}[\textit{Stage 1a trans. ``calm''}]\\
    Membingitkan& \bigsq & \bigsq & \bigsq & \bigsq & \bigsq \\
    {}[\textit{Stage 1a trans. ``annoying''}]\\
    Meriah& \bigsq & \bigsq & \bigsq & \bigsq & \bigsq\\
    {}[\textit{Stage 1a trans. ``eventful''}]\\
    Membosankan & \bigsq & \bigsq & \bigsq & \bigsq & \bigsq\\ 
    {}[\textit{Stage 1a trans. ``monotonous''}]\\
\bottomrule
\end{tabularx}
\end{sffamily}

\newcommand{\linecirc}{
\begin{tikzpicture}
        \draw [color=black, fill=black, radius=.1]
          (1, 0) circle[]
          (12, 0) circle[];
        \draw [-] (1, 0) -- (12, 0);
\end{tikzpicture}
}

\begin{sffamily}
\setcounter{table}{2}
\small
\noindent\textbf{\color{scolor}Table \thetable}\\
\normalfont\sffamily The translated PAQ questionnaire to Bahasa Melayu, as used in Stage 1c. The upright text shows the Bahasa Melayu version as presented to the participants. The following italicized text in square brackets indicates the corresponding English counterpart. The attributes were rated on a visual analog scale (VAS) from ``Not at all'' to ``Completely''.
\label{tab:stage1cqn}
\small
\setlength{\tabcolsep}{0pt}
\vskip1em\noindent
\begin{tabularx}{\textwidth}{%
    l
    @{\hspace{-2.5em}}*{5}{>{\centering\arraybackslash}X}%
}
\toprule
    \multicolumn{6}{p{17cm}}{%
        Secara keseluruhannya, bagaimanakah anda menilai bunyi persekitaran anda sekarang? \newline
        [\textit{Overall, how would you describe the present surrounding sound environment?}] 
    } \\
\midrule
    & Tidak sama sekali           
    &             
    &    
    &        
    & Melampau\\ 
    & [\textit{Not at all}]                      
    &              
    &        
    &             
    & [\textit{Completely}]   \\ 
\midrule
    1. Meriah [\textit{``eventful''}]& \multicolumn{5}{c}{\linecirc}\\
    \\
    2. Bersemarak [\textit{``vibrant''}]& \multicolumn{5}{c}{\linecirc}\\
    \\
    3. Menyenangkan [\textit{``pleasant''}]& \multicolumn{5}{c}{\linecirc}\\
    \\
    4. Tenang [\textit{``calm''}]& \multicolumn{5}{c}{\linecirc}\\
    \\
    5. Tidak meriah [\textit{``uneventful''}]& \multicolumn{5}{c}{\linecirc}\\
    \\
    6. Kurang kepelbagaian \\dan membosankan 
    [\textit{``monotonous''}]& 
    \multicolumn{5}{c}{\linecirc}\\
    \\
    7. Membingitkan [\textit{``annoying''}]& \multicolumn{5}{c}{\linecirc}\\
    \\
    8. Kacau-bilau [\textit{``chaotic''}]& \multicolumn{5}{c}{\linecirc}\\
\bottomrule
\end{tabularx}
\end{sffamily}

\begin{sffamily}
\setcounter{table}{3}
\small
\noindent\textbf{\color{scolor}Table \thetable}\\
\normalfont\sffamily The translated supplementary PAQ questionnaire to Bahasa Melayu with other translations, as used in Stage 1c. This questionnaire was provided to the participants after completion of the experiential evaluation with the questionnaire in Table B.2. The upright text shows the Bahasa Melayu version along with italicized text in square brackets indicating the corresponding English counterpart, both of which was presented to the participants for further comments.
\label{tab:stage1csup}
\small
\setlength{\tabcolsep}{0pt}
\vskip1em\noindent
\begin{tabularx}{\textwidth}{%
    l
    @{\hspace{-2.5em}}*{5}{>{\centering\arraybackslash}X}%
}
\toprule
    \multicolumn{6}{p{17cm}}{%
        Secara keseluruhannya, bagaimanakah anda menilai bunyi persekitaran anda sekarang? \newline
        [\textit{Overall, how would you describe the present surrounding sound environment?}] 
    } \\
\midrule
    & Tidak sama sekali           
    &             
    &    
    &        
    & Melampau\\ 
    & [\textit{Not at all}]                      
    &              
    &        
    &             
    & [\textit{Completely}]   \\ 
\midrule
    \\
    Meriah [\textit{``eventful''}]& \multicolumn{5}{c}{\linecirc}\\
    \\
    Bersemarak; Rancak [\textit{``vibrant''}]& \multicolumn{5}{c}{\linecirc}\\
    \\
    Menyenangkan [\textit{``pleasant''}] & \multicolumn{5}{c}{\linecirc}\\
    \\
    Tenang [\textit{``calm''}]& \multicolumn{5}{c}{\linecirc}\\
    \\
    Tidak meriah; \\Hambar dan senyap  [\textit{``uneventful''}] & \multicolumn{5}{c}{\linecirc}\\
    \\
    Kurang kepelbagaian \\dan membosankan; \\
    Membosankan [\textit{``monotonous''}]& \multicolumn{5}{c}{\linecirc}\\
    \\
    Membingitkan; \\Menjengkelkan [\textit{``annoying''}]& \multicolumn{5}{c}{\linecirc}\\
    \\
    Kacau-bilau [\textit{``chaotic''}] & \multicolumn{5}{c}{\linecirc}\\
    \\

\bottomrule
\end{tabularx}
\end{sffamily}

\setcounter{table}{1}
\renewcommand{\thetable}{\Alph{section}.\arabic{table}}

\section{Results of Statistical Tests on the Evaluation Scores}
\setcounter{table}{0}
\begin{sffamily}
\begin{nolinenumbers}
\small%
\refstepcounter{table}\label{tab:main_scores}
\noindent\textbf{\color{scolor}Table \thetable}\par%
\noindent{Mean evaluation scores for the attributes across the combined, Malaysian-only, and Singaporean-only populations.}%
\end{nolinenumbers}
\small%
\begin{longtable}{p{2.7cm}p{1.5cm}p{6cm}*{3}{>{\centering\arraybackslash}p{1.6cm}}}
\toprule
&&& \multicolumn{3}{c}{Mean Score}\\
\cmidrule{4-6}
Attribute & Criterion &  Translation Candidate & Combined & MY & SG\\
\midrule
\endhead

    \\
    \multicolumn{6}{r}{[Continued on next page]} \\
\endfoot
\endlastfoot

pleasant & APPR & menyenangkan &  0.840 &    0.857 &     0.824 \\*
        \cmidrule{2-6}   & UNDR & menyenangkan &  0.910 &    0.900 &     0.918 \\*
        \cmidrule{2-6}   & CLAR & menyenangkan &  0.526 &    0.557 &     0.498 \\*
         \cmidrule{2-6}  & ORTH & menyenangkan &  0.613 &    0.633 &     0.594 \\*
        \cmidrule{2-6}   & ANTO & menyenangkan &  0.624 &    0.647 &     0.603 \\*
        \cmidrule{2-6}   & NCON & menyenangkan &  0.444 &    0.483 &     0.408 \\*
        \cmidrule{2-6}   & IBAL & menyenangkan &  0.563 &    0.633 &     0.500 \\
        \midrule
annoying & APPR & membingitkan &  0.803 &    0.783 &     0.821 \\
           &      & menjengkelkan &  0.846 &    0.897 &     0.800 \\
           \cmidrule{2-6}
           & UNDR & membingitkan &  0.773 &    0.767 &     0.779 \\
           &      & menjengkelkan &  0.711 &    0.823 &     0.609 \\
           \cmidrule{2-6} \pagebreak
           & CLAR & membingitkan &  0.608 &    0.608 &     0.608 \\
           &      & menjengkelkan &  0.648 &    0.673 &     0.626 \\
           \cmidrule{2-6}
           & ORTH & membingitkan &  0.590 &    0.647 &     0.539 \\
           &      & menjengkelkan &  0.530 &    0.540 &     0.521 \\
           \cmidrule{2-6}
           & ANTO & membingitkan &  0.716 &    0.730 &     0.703 \\
           &      & menjengkelkan &  0.740 &    0.753 &     0.727 \\
           \cmidrule{2-6}
           & NCON & membingitkan &  0.554 &    0.543 &     0.564 \\
           &      & menjengkelkan &  0.649 &    0.675 &     0.626 \\
           \cmidrule{2-6}
           & IBAL & membingitkan &  0.565 &    0.633 &     0.503 \\
           &      & menjengkelkan &  0.756 &    0.777 &     0.736 \\
           \midrule 
eventful & APPR & meriah &  0.673 &    0.677 &     0.670 \\*
         \cmidrule{2-6}  & UNDR & meriah &  0.929 &    0.960 &     0.900 \\*
          \cmidrule{2-6} & CLAR & meriah &  0.540 &    0.585 &     0.500 \\*
         \cmidrule{2-6}  & ORTH & meriah &  0.416 &    0.467 &     0.370 \\*
         \cmidrule{2-6}  & ANTO & meriah &  0.586 &    0.603 &     0.570 \\*
         \cmidrule{2-6}  & NCON & meriah &  0.406 &    0.452 &     0.364 \\*
        \cmidrule{2-6}   & IBAL & meriah &  0.608 &    0.570 &     0.642 \\
        \midrule
uneventful & APPR & tidak meriah &  0.713 &    0.737 &     0.691 \\
       \cmidrule{2-6}    & UNDR & tidak meriah &  0.902 &    0.920 &     0.885 \\
        \cmidrule{2-6}   & CLAR & tidak meriah &  0.579 &    0.630 &     0.532 \\
        \cmidrule{2-6}   & ORTH & tidak meriah &  0.695 &    0.687 &     0.703 \\
        \cmidrule{2-6}   & ANTO & tidak meriah &  0.668 &    0.677 &     0.661 \\
        \cmidrule{2-6}   & NCON & tidak meriah &  0.513 &    0.532 &     0.497 \\
         \cmidrule{2-6}  & IBAL & tidak meriah &  0.684 &    0.683 &     0.685 \\
         \midrule
calm & APPR & menenangkan &  0.838 &    0.817 &     0.858 \\
        &      & tenang &  0.968 &    0.953 &     0.982 \\
        \cmidrule{2-6}
        & UNDR & menenangkan &  0.910 &    0.910 &     0.909 \\
        &      & tenang &  0.979 &    0.963 &     0.994 \\
        \cmidrule{2-6}
        & CLAR & menenangkan &  0.482 &    0.528 &     0.439 \\
        &      & tenang &  0.514 &    0.510 &     0.518 \\
        \cmidrule{2-6}
        & CONN & menenangkan &  0.589 &    0.537 &     0.636 \\
        &      & tenang &  0.613 &    0.597 &     0.629 \\
        \cmidrule{2-6}
        & IBAL & menenangkan &  0.584 &    0.533 &     0.630 \\
        &      & tenang &  0.586 &    0.580 &     0.591 \\
        \midrule
chaotic & APPR & huru-hara &  0.946 &    0.973 &     0.921 \\
        &      & kelam-kabut &  0.790 &    0.880 &     0.709 \\
        \cmidrule{2-6}
        & UNDR & huru-hara &  0.954 &    0.967 &     0.942 \\
        &      & kelam-kabut &  0.948 &    0.970 &     0.927 \\
        \cmidrule{2-6}
        & CLAR & huru-hara &  0.614 &    0.715 &     0.523 \\
        &      & kelam-kabut &  0.655 &    0.683 &     0.629 \\
        \cmidrule{2-6}
        & CONN & huru-hara &  0.467 &    0.380 &     0.547 \\
        &      & kelam-kabut &  0.414 &    0.367 &     0.458 \\
        \cmidrule{2-6}
        & IBAL & huru-hara &  0.713 &    0.747 &     0.682 \\
        &      & kelam-kabut &  0.708 &    0.733 &     0.685 \\
        \midrule \pagebreak
monotonous & APPR & kurang kepelbagaian oleh itu membosankan &  0.776 &    0.787 &     0.767 \\
        &      & membosankan &  0.765 &    0.750 &     0.779 \\
        &      & tidak berubah oleh itu membosankan &  0.654 &    0.617 &     0.688 \\
        \cmidrule{2-6}
        & UNDR & kurang kepelbagaian oleh itu membosankan &  0.805 &    0.833 &     0.779 \\
        &      & membosankan &  0.946 &    0.957 &     0.936 \\
        &      & tidak berubah oleh itu membosankan &  0.765 &    0.753 &     0.776 \\
        \cmidrule{2-6}
        & CLAR & kurang kepelbagaian oleh itu membosankan &  0.537 &    0.543 &     0.530 \\
        &      & membosankan &  0.549 &    0.578 &     0.523 \\
        &      & tidak berubah oleh itu membosankan &  0.564 &    0.573 &     0.556 \\
        \cmidrule{2-6}
        & CONN & kurang kepelbagaian oleh itu membosankan &  0.486 &    0.462 &     0.508 \\*
        &      & membosankan &  0.496 &    0.493 &     0.498 \\*
        &      & tidak berubah oleh itu membosankan &  0.454 &    0.438 &     0.468 \\
        \cmidrule{2-6}
        & IBAL & kurang kepelbagaian oleh itu membosankan &  0.562 &    0.643 &     0.488 \\*
        &      & membosankan &  0.529 &    0.540 &     0.518 \\*
        &      & tidak berubah oleh itu membosankan &  0.546 &    0.597 &     0.500 \\
        \midrule
vibrant & APPR & bersemarak &  0.790 &    0.743 &     0.833 \\
        &      & rancak &  0.722 &    0.707 &     0.736 \\
        \cmidrule{2-6}
        & UNDR & bersemarak &  0.689 &    0.737 &     0.645 \\
        &      & rancak &  0.797 &    0.853 &     0.745 \\
        \cmidrule{2-6}
        & CLAR & bersemarak &  0.456 &    0.508 &     0.408 \\
        &      & rancak &  0.475 &    0.507 &     0.447 \\
        \cmidrule{2-6}
        & CONN & bersemarak &  0.623 &    0.607 &     0.638 \\
        &      & rancak &  0.607 &    0.557 &     0.653 \\
        \cmidrule{2-6}
        & IBAL & bersemarak &  0.744 &    0.760 &     0.730 \\
        &      & rancak &  0.716 &    0.680 &     0.748 \\
\bottomrule
    
    \end{longtable}
\end{sffamily}
\vfill
\begin{sffamily}
\begin{nolinenumbers}
\small\par
\vspace{\baselineskip}
\addtocounter{table}{-1}\refstepcounter{table}\label{tab:prentice}
\noindent\textbf{\color{scolor}Table \thetable}\par%
\noindent{Results of the cross-national tests for differences in distributions of the evaluation scores, using either the Prentice test or the Kruskal-Wallis test. Double asterisks (**) and single asterisk (*) indicate statistical significance at \SI{1}{\percent} and \SI{5}{\percent}, respectively.}%
\end{nolinenumbers}
\small
\vspace{0.5\baselineskip}
\noindent\begin{tabularx}{\columnwidth}{ll*{7}{>{\raggedleft\arraybackslash}X}}
\toprule
&& \multicolumn{7}{c}{Criterion}\\
\cmidrule(lr){3-9}
Attribute &    Test & APPR &             UNDR &              CLAR &     ORTH &     ANTO &             NCON &     IBAL \\
\midrule
pleasant   &  Kruskal-Wallis &           {0.352} &          {0.822} &           {0.178} &  {0.581} &  {0.656}                &  {0.051} &  {0.089} \\
annoying   &   Prentice &           {0.225} &  *\textbf{0.034} &           {0.784} &  {0.296} &  {0.302}                &  {0.872} &  {0.105} \\
eventful   &  Kruskal-Wallis &           {0.588} &  *\textbf{0.041} &           {0.092} &  {0.196} &  {0.383}                &  {0.065} &  {0.288} \\
uneventful &  Kruskal-Wallis &           {0.409} &          {0.116} &           {0.157} &  {0.691} &  {0.702}                &  {0.499} &  {0.956} \\
\bottomrule\\
\toprule
&& \multicolumn{7}{c}{Criterion}\\
\cmidrule(lr){3-9}
Attribute &    Test &          APPR &             UNDR &              CLAR &     &&          CONN & IBAL \\
\midrule
calm       &   Prentice &           {0.212} &          {0.243} &           {0.324} &       &       &  *\textbf{0.026}        &  {0.512} \\
chaotic    &   Prentice &  **\textbf{0.007} &          {0.153} &  **\textbf{0.007} &       &       &  *\textbf{0.011}        &  {0.210} \\
vibrant    &   Prentice &           {0.251} &          {0.054} &   *\textbf{0.037} &       &       &          {0.086} &         {0.897}\\
monotonous &   Prentice &           {0.577} &          {0.979} &           {0.714} &       &       &          {0.424} &         {0.057} \\
\bottomrule
    \end{tabularx}
    
\end{sffamily}
\vfill
\begin{table}[t]
    \centering
\caption{Results of the posthoc Mann–Whitney–Wilcoxon pairwise tests for cross-national differences in distributions of the evaluation scores, for evaluation criteria with significant differences from the omnibus tests. Double asterisks (**) and single asterisk (*) indicate statistical significance at \SI{1}{\percent} and \SI{5}{\percent}, respectively.}
    \begin{tabular}{lllrrr}
    \toprule
    &&&& \multicolumn{2}{c}{Mean Score}\\
    \cmidrule(lr){5-6}
    Attribute & Criterion & Translation Candidate & {\textit{p}-value} & \multicolumn{1}{c}{MY} & \multicolumn{1}{c}{SG}\\
    \midrule
    annoying & UNDR & membingitkan &           {1.000} &    0.767 &     0.779 \\
            &      & menjengkelkan &  **\textbf{0.003} &    \textbf{0.823} &     0.609\\
            \midrule
    vibrant & CLAR & rancak &           {0.629} &    0.507 &     0.447\\
            &      & bersemarak &           {0.109} &    0.508 &     0.408 \\
            \midrule
    calm & CONN & tenang &           {0.771} &    0.597 &     0.629 \\
            &      & menenangkan &   *\textbf{0.048} &    0.537 &     \textbf{0.636} \\
            \midrule
    chaotic & APPR & huru-hara &           {0.384} &    0.973 &     0.921\\
            &      & kelam-kabut &   *\textbf{0.032} &    \textbf{0.880} &     0.709 \\
            \cmidrule{2-6}
            & CLAR & huru-hara &  **\textbf{0.006} &   \textbf{ 0.715 }&     0.523 \\
            &      & kelam-kabut &           {0.762} &    0.683 &     0.629\\
            \cmidrule{2-6}
            & CONN & huru-hara &   *\textbf{0.043} &    0.380 &     \textbf{0.547} \\
            &      & kelam-kabut &           {0.404}&    0.367 &     0.458 \\
    \bottomrule
    \end{tabular}
    \label{tab:wilcoxon}
\end{table}

\FloatBarrier
\begin{sffamily}
\begin{nolinenumbers}
\small%
\refstepcounter{table}
\noindent\textbf{\color{scolor}Table \thetable}\par%
\noindent{$p$-values of the Kruskal--Wallis tests and the posthoc Conover--Iman tests. Double asterisks (**) and single asterisk (*) indicate statistical significance at \SI{1}{\percent} and \SI{5}{\percent}, respectively.%
}%
\end{nolinenumbers}
\small%
\begin{longtable}{lll>{\raggedright}p{10.5cm}r}
    \toprule
    Attribute &
    Crit. &
    CCR &
    Test &
    $p$-value\\
    \midrule
\endhead

    \\
    \multicolumn{5}{r}{[Continued on next page]} \\
\endfoot

    \bottomrule
    \label{tab:citp}
\endlastfoot
annoying & APPR & MY & Kruskal-Wallis &    *\textbf{0.028} \\*
           &      & SG & Kruskal-Wallis &            {0.757} \\
\cmidrule{2-5}
           & UNDR & MY & Kruskal-Wallis &            {0.441} \\*
           &      & SG & Kruskal-Wallis &   **\textbf{0.009} \\
\cmidrule{2-5}
           & CLAR & MY & Kruskal-Wallis &            {0.199} \\*
           &      & SG & Kruskal-Wallis &            {0.455} \\
\cmidrule{2-5}
           & ORTH & MY & Kruskal-Wallis &            {0.261} \\*
           &      & SG & Kruskal-Wallis &            {0.803} \\
\cmidrule{2-5}
           & ANTO & MY & Kruskal-Wallis &            {0.577} \\*
           &      & SG & Kruskal-Wallis &            {0.979} \\
\cmidrule{2-5}
           & NCON & MY & Kruskal-Wallis &    *\textbf{0.026} \\*
           &      & SG & Kruskal-Wallis &            {0.315} \\
\cmidrule{2-5}
           & IBAL & MY & Kruskal-Wallis &            {0.081} \\*
           &      & SG & Kruskal-Wallis &   **\textbf{0.003} \\
\midrule
calm & APPR & MY & Kruskal-Wallis &   **\textbf{0.001} \\*
           &      & SG & Kruskal-Wallis &  **\textbf{0.0004} \\
\cmidrule{2-5}
           & UNDR & MY & Kruskal-Wallis &    *\textbf{0.014} \\*
           &      & SG & Kruskal-Wallis &  **\textbf{0.0002} \\
\cmidrule{2-5}
           & CLAR & MY & Kruskal-Wallis &            {0.846} \\*
           &      & SG & Kruskal-Wallis &            {0.231} \\
\cmidrule{2-5}
           & CONN & MY & Kruskal-Wallis &            {0.594} \\*
           &      & SG & Kruskal-Wallis &            {0.908} \\
\cmidrule{2-5}
           & IBAL & MY & Kruskal-Wallis &            {0.622} \\*
           &      & SG & Kruskal-Wallis &            {0.531} \\
\midrule
chaotic & APPR & MY & Kruskal-Wallis &   **\textbf{0.007} \\*
           &      & SG & Kruskal-Wallis &  **\textbf{0.0004} \\
\cmidrule{2-5}
           & UNDR & MY & Kruskal-Wallis &            {0.944} \\*
           &      & SG & Kruskal-Wallis &            {0.884} \\
\cmidrule{2-5}
           & CLAR & MY & Kruskal-Wallis &            {0.737} \\*
           &      & SG & Kruskal-Wallis &            {0.085} \\
\cmidrule{2-5}
           & CONN & MY & Kruskal-Wallis &            {0.864} \\*
           &      & SG & Kruskal-Wallis &            {0.181} \\
\cmidrule{2-5}
           & IBAL & MY & Kruskal-Wallis &            {0.581} \\*
           &      & SG & Kruskal-Wallis &            {0.770} \\
\midrule
vibrant & APPR & MY & Kruskal-Wallis &            {0.580} \\*
           &      & SG & Kruskal-Wallis &            {0.314} \\
\cmidrule{2-5}
           & UNDR & MY & Kruskal-Wallis &    *\textbf{0.028} \\*
           &      & SG & Kruskal-Wallis &            {0.146} \\
\cmidrule{2-5}
           & CLAR & MY & Kruskal-Wallis &            {0.882} \\*
           &      & SG & Kruskal-Wallis &            {0.545} \\
\cmidrule{2-5}
           & CONN & MY & Kruskal-Wallis &            {0.278} \\*
           &      & SG & Kruskal-Wallis &            {0.969} \\
\cmidrule{2-5}
           & IBAL & MY & Kruskal-Wallis &            {0.311} \\*
           &      & SG & Kruskal-Wallis &            {0.875} \\
\midrule
monotonous & APPR & MY & Kruskal-Wallis &    *\textbf{0.014} \\
\cmidrule{4-5}
& & & Conover-Iman \hfill (o/i: oleh itu) & \\*
           &      &    & kurang kepelbagaian o/i membosankan - membosankan &            {1.000} \\*
           &      &    & kurang kepelbagaian o/i membosankan - tidak berubah o/i membosankan &   **\textbf{0.009} \\*
           &      &    & membosankan - tidak berubah o/i membosankan &    *\textbf{0.029} \\*
\cmidrule{3-5}           &      & SG & Kruskal-Wallis &            {0.312} \\
\cmidrule{2-5}
           & UNDR & MY & Kruskal-Wallis &  **\textbf{0.0001} \\
           \cmidrule{4-5}
& & & Conover-Iman & \\*
           &      &    & kurang kepelbagaian o/i membosankan - membosankan &  **\textbf{0.0001} \\*
           &      &    & kurang kepelbagaian o/i membosankan - tidak berubah o/i membosankan &            {0.271} \\*
           &      &    & membosankan - tidak berubah o/i membosankan &   **\textbf{3e-07} \\*
\cmidrule{3-5}           &      & SG & Kruskal-Wallis &   **\textbf{0.003} \\
           \cmidrule{4-5}
& & & Conover-Iman & \\*
           &      &    & kurang kepelbagaian o/i membosankan - membosankan &   **\textbf{0.007} \\*
           &      &    & kurang kepelbagaian o/i membosankan - tidak berubah o/i membosankan &            {1.000} \\*
           &      &    & membosankan - tidak berubah o/i membosankan &   **\textbf{0.003} \\
\cmidrule{2-5}
           & CLAR & MY & Kruskal-Wallis &            {0.787} \\*
           &      & SG & Kruskal-Wallis &            {0.889} \\
\cmidrule{2-5}
           & CONN & MY & Kruskal-Wallis &            {0.469} \\*
           &      & SG & Kruskal-Wallis &            {0.831} \\
\cmidrule{2-5}
           & IBAL & MY & Kruskal-Wallis &            {0.409} \\*
           &      & SG & Kruskal-Wallis &            {0.761} \\
\end{longtable}
\end{sffamily}



\end{document}